\documentclass[aps,prb,twocolumn,superscriptaddress,floatfix]{revtex4-2}
\usepackage{graphicx}
\usepackage{amsmath,amssymb}
\usepackage[hidelinks]{hyperref}
\usepackage{booktabs}
\usepackage{xcolor}

\begin{document}

\title{Spin fluctuations and nearly isotropic quasiparticles in the doped three-dimensional Hubbard model}

\author{Pengcheng Hou}
\affiliation{Hefei National Laboratory, University of Science and Technology of China, Hefei 230088, China}

\author{Kun Chen}
\affiliation{CAS Key Laboratory of Theoretical Physics, Institute of Theoretical Physics, Chinese Academy of Sciences, Beijing 100190, China}

\author{Youjin Deng}
\email{yjdeng@ustc.edu.cn}
\affiliation{Department of Modern Physics, University of Science and Technology of China, Hefei, Anhui 230026, China}
\affiliation{Hefei National Laboratory, University of Science and Technology of China, Hefei 230088, China}

\date{\today}

\begin{abstract}
    Away from half filling and above its N\'eel temperature, the three-dimensional Hubbard model is a paramagnetic metal with antiferromagnetic spin fluctuations, for which controlled momentum-resolved results have been lacking.
    Using connected-determinant diagrammatic Monte Carlo, we compute the momentum-resolved self-energy for $n = 0.4$--$0.95$, $T/t = 0.05$--$0.33$ and $U/t = 4$--$8$, and the spin susceptibility over the Brillouin zone at twelve of these points, focusing on $U/t = 4$.
    At $U/t=4$ and toward half filling the spin fluctuations become strong: for $n = 0.7$--$0.9$ at $T/t=0.14$ they enhance the susceptibility peak up to sevenfold and raise the momentum-dependent part of the scattering rate to four to five times its second-order value.
    They are driven by an effective spin interaction screened to about two thirds of $U$ and nearly uniform over the Brillouin zone.
    Yet the peak height is set by the slightly smaller value of this interaction at the peak wave vector, and near magnetic order even this small difference changes the height by a large factor.
    The two-particle self-consistent theory, with a single constant interaction, reproduces the peak because its sum rule fixes that constant close to this value, but for $n = 0.7$--$0.9$ at $T/t = 0.14$ the self-energy it implies overestimates the local part of the scattering rate and underestimates its momentum-dependent part.
    Despite their strength, the fluctuations are short ranged, with a correlation length of at most about two lattice spacings, and the self-energy depends almost only on the band energy: for $n = 0.7$--$0.9$ and $T/t \geq 0.14$ the scattering rate near the Fermi surface varies by at most about $1\%$ at fixed band energy, without hot spots.
    At $U/t = 4$ dynamical mean-field theory (DMFT) reproduces the local scattering rate and quasiparticle weight within a few percent, and the nonlocal part it misses is, to good approximation, a function of the band energy.
    Strong spin fluctuations thus coexist with nearly isotropic quasiparticles, and these momentum-resolved results, with their error budgets, provide a benchmark for extensions of DMFT and for spin-fluctuation theories.
\end{abstract}

\maketitle

\section{Introduction}
\label{sec:intro}

The Hubbard model on the simple-cubic lattice is the minimal model of interacting electrons in three dimensions.
At half filling it is well characterized: its N\'eel temperature is known over the whole coupling range~\cite{Staudt2000,Kent2005,Kozik2013,Rohringer2011,Hirschmeier2015,SunXu2025,Song2025PRB}, the ordered phase has been reached by large-order diagrammatic expansions~\cite{Garioud2024}, and fermionic atoms in optical lattices have progressed from antiferromagnetic correlations~\cite{Hart2015} to the N\'eel transition itself~\cite{Shao2024}.
Away from half filling much less is established.
Controlled methods have computed the thermodynamics of the doped model, including its entropy, down to moderate temperature~\cite{Kozik2010,Khatami2016,SongDengHe2026}.
Diagrammatic Monte Carlo has mapped its N\'eel boundary near half filling together with the equal-time spin structure factor~\cite{Lenihan2022}, and dynamical mean-field theory (DMFT) predicts this boundary over a wide range of doping and coupling, including incommensurate spiral phases~\cite{Rampon2025}.
Above the boundary the doped model is a paramagnetic metal: itinerant quasiparticles coexist with antiferromagnetic spin fluctuations, which doping moves from $(\pi,\pi,\pi)$ along the nesting ridge $(\pi,\pi,q_z)$, where the bare particle-hole bubble is largest~\cite{TahvildarZadeh1997,Schafer2017,Lenihan2022}.

Two questions decide what these fluctuations do to the electrons: how strong they are, and how far their imprint on the quasiparticles reaches.
In two dimensions the imprint is pronounced.
Near half filling and at low temperature the spin correlation length grows exponentially, and the fluctuations produce \emph{hot spots}, points on the Fermi surface with strongly enhanced scattering, and eventually a pseudogap~\cite{Vilk1997,Simkovic2020,Wu2017,Gunnarsson2015}; at the Fermi level of the half-filled model the self-energy then depends on the direction of the momentum, not only on the band energy~\cite{Pudleiner2016}.
In three dimensions at half filling, cluster calculations find nonlocal spectral features that DMFT lacks~\cite{Fuchs2011}, the dynamical vertex approximation (D$\Gamma$A) finds the nonlocal, i.e., momentum-dependent, part of the self-energy important mainly close to the N\'eel temperature~\cite{Rohringer2011,Rohringer2016}, and auxiliary-field quantum Monte Carlo finds the same quasiparticle weight, within error bars, at all Fermi momenta~\cite{Song2025PRB}.
In the doped model at larger coupling, D$\Gamma$A finds the quasiparticle weight nearly momentum independent while the static real part of the self-energy is strongly nonlocal~\cite{SchaferTomczak2015}, and a cluster calculation reports a pseudogap closer to half filling~\cite{Kellar2023}.
Both calculations of the doped model approximate the two-particle vertex, the effective interaction between two electrons in the presence of all others, as local or confined to a small cluster, so neither determines how strong the fluctuations are or how far their imprint reaches without that approximation.

Each question has a natural yardstick, and each touches a simplification made by the practical theories of the doped metal.
The strength of the fluctuations is set by the effective interaction that drives them.
In the random-phase approximation (RPA) this is the bare $U$.
Particle-particle scattering screens it, the Kanamori--Brueckner mechanism by which two electrons lower their repulsion by avoiding each other~\cite{Kanamori1963}.
Close to the Stoner instability even a small change of the effective interaction changes the enhancement of the response by a large factor.
The two-particle self-consistent (TPSC) theory keeps one constant effective interaction per channel, fixed by local sum rules that tie the fluctuations to the double occupancy, which it estimates with an ansatz~\cite{Vilk1994,Vilk1997,Tremblay2012}; it has recently been combined with DMFT~\cite{Martin2023}.

The imprint is the momentum dependence of the self-energy, whose weak-coupling reference is second-order perturbation theory, and the question is whether it follows the band energy or singles out hot spots.
DMFT discards this momentum dependence by making the irreducible vertex, and hence the self-energy, local~\cite{Georges1996,Pruschke1995}.
Locality is exact in infinite dimensions and expected to be accurate in three~\cite{Georges1996,Rohringer2018}, so DMFT underlies realistic calculations for three-dimensional materials, with nonlocal corrections added through $GW$~\cite{Tomczak2014} or through diagrammatic extensions such as D$\Gamma$A~\cite{Toschi2007,Katanin2009,Rohringer2018,Antipov2014,WangHaule2025,Leusch2026}.
In two dimensions controlled references for both quantities, the effective spin interaction and the momentum-resolved self-energy, exist~\cite{Simkovic2022,Kim2020,Simkovic2020,Wu2017}: effective vertices have been extracted from exact susceptibilities~\cite{Bulut1993,Vilk1994}, approximate theories have been tested against exact, quantum-simulator and multi-method data~\cite{Schaefer2021,HuangWu2024,LeBlanc2015}, and improved versions of TPSC have been benchmarked in the same way~\cite{GauvinNdiaye2023}.
To our knowledge, in three dimensions TPSC has been compared with controlled data only at half filling~\cite{DareAlbinet2000,DelRe2025}, and no controlled calculation has given the self-energy or the dynamical susceptibility of the doped model.

Here we provide such a reference: the momentum-resolved self-energy together with the dynamical spin and charge susceptibilities over the whole Brillouin zone, at the same parameters and with error budgets.
We use the connected-determinant (CDet) diagrammatic Monte Carlo method~\cite{Rossi2017,Moutenet2018,SimkovicKozik2019} in its fixed-density form~\cite{Simkovic2022,Simkovic2024Science}, which samples all Feynman diagrams of the self-energy and the susceptibilities up to sixth to tenth order in $U$ and resums the series, with no approximation for the vertex.
On a $20^3$ lattice we compute the self-energy $\Sigma(\mathbf k, i\omega_n)$ at $48$ filling--temperature points, $n = 0.4$--$0.95$ at $T/t = 0.05$--$0.33$, and the spin and charge susceptibilities at twelve of them with $n = 0.4$--$0.9$.
We focus on $U/t = 4$, a third of the bandwidth, where the resummed series are well controlled; $U/t = 6$ and $8$ extend the local results, qualitatively near half filling at $U/t=8$, and show the trends of the nonlocal ones.
Writing each measured spin susceptibility in the RPA form $\chi_\mathrm{sp} = \chi_0/(1 - U_\mathrm{sp}\chi_0/2)$, with $\chi_0$ the bare bubble of the same correlator, defines an effective spin interaction, which we call the \emph{spin vertex} $U_\mathrm{sp}(\mathbf q, i\Omega_m)$, at every momentum and at the four lowest bosonic frequencies.
It is the value that a theory with one constant vertex would need in order to reproduce the response there.
Comparing it with $U$ and across the zone shows how far a single constant interaction, as in TPSC, can describe the response, and comparison with single-site DMFT shows how well a local theory captures the local self-energy.
Inserting $U_\mathrm{sp}$ into the exchange of spin fluctuations shows how much of the momentum dependence of the self-energy the fluctuations account for, and exact low-order diagrams on the same lattice identify the scattering channels.

We find that the spin fluctuations of the doped metal are strong but short ranged.
They are strong even at $U/t=4$: at $n = 0.9$ and $T/t = 0.14$ the spin susceptibility at its peak is enhanced sevenfold over the bare bubble, and for $n = 0.7$--$0.9$ on the same isotherm the momentum dependence of the scattering rate reaches $4.0$--$5.4$ times its second-order value.
Their strength is set by the spin vertex, which is screened to about two thirds of $U$, at second order mainly by particle-particle scattering, and varies little over the zone.
Its slightly smaller value on the nesting ridge sets the height of the peak, and TPSC reproduces the peak because the sum rule places the TPSC constant close to this value.
The spin fluctuations are nevertheless short ranged: their correlation length is at most about two lattice spacings, and for $n = 0.7$--$0.9$ and $T/t \geq 0.14$ the self-energy is almost a function of the band energy, without hot spots.
Single-site DMFT reproduces the local scattering rate and quasiparticle weight to within a few percent, so a DMFT extension at this coupling has to add a nonlocal self-energy that depends, to good approximation, on the band energy alone.
For $n = 0.7$--$0.9$ on the $T/t=0.14$ isotherm the static exchange of the measured fluctuations follows the growth of the nonlocal scattering rate, and exact low-order diagrams attribute the growth to spin-fluctuation diagrams rather than to the particle-particle ladder.
Exchange through a bare fermion-boson vertex, the coupling of an electron to a fluctuation, nevertheless overestimates the local scattering and still falls short of the nonlocal part, both with the measured fluctuations and in TPSC. No constant rescaling of that vertex removes the imbalance.
The momentum-resolved self-energies and vertices provide a benchmark for such theories.

Section~\ref{sec:method} defines the model, the method and the observables; Sec.~\ref{sec:vertex} presents the spin vertex and the spin-fluctuation peak, Sec.~\ref{sec:sigma} the self-energy, and Sec.~\ref{sec:discussion} the consequences for approximate theories.
Appendices~\ref{app:resum} and~\ref{app:checks} give the error budget and the main checks, and Appendices~\ref{app:vertices} and~\ref{app:sigma} the analyses behind Secs.~\ref{sec:vertex} and~\ref{sec:sigma}.

\section{Model, method and definitions}
\label{sec:method}

\subsection{Model and parameter set}
\label{sec:model}

We study the single-band Hubbard model on the simple-cubic lattice,
\begin{equation}
    \hat{H} = -t\!\sum_{\langle ij\rangle,\sigma} c_{i\sigma}^\dagger c_{j\sigma}
    + U\sum_i n_{i\uparrow} n_{i\downarrow}
    - \mu \sum_{i,\sigma} n_{i\sigma},
    \label{eq:hamiltonian}
\end{equation}
with nearest-neighbor hopping $t=1$ as the unit of energy, band energy $\varepsilon_\mathbf{k} = -2t\sum_d \cos k_d$, bandwidth $W=12t$ and inverse temperature $\beta = 1/T$.
All production runs evaluate the diagrams with the propagators of a periodic lattice of linear size $L=20$ ($286$ inequivalent momenta under $O_h$ symmetry).

\begin{figure}[t]
    \centering
    \includegraphics[width=\columnwidth]{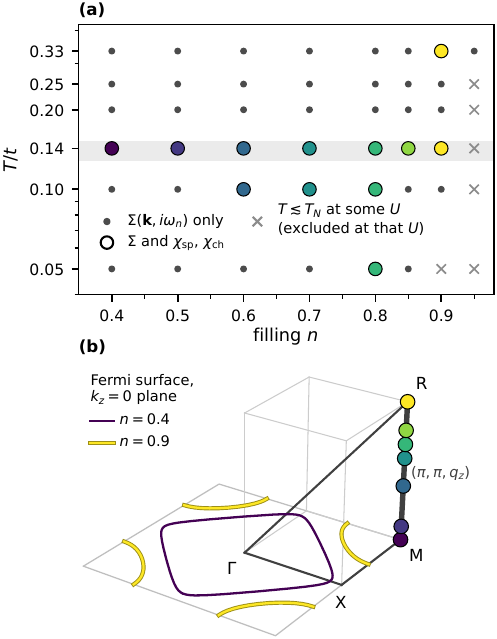}
    \caption{The parameter set covers the doped paramagnet, and with increasing filling the spin-fluctuation peak moves along the nesting ridge from $M$ to $R$.
        (a) Filling--temperature plane, logarithmic in $T$; the momentum-resolved self-energy is computed at all $48$ points.
        Small dots: self-energy only; large circles: the twelve points that also carry the spin and charge susceptibilities, colored by filling, on the one scale used wherever color encodes filling; crosses: points at or beyond the magnetic boundary of Ref.~\cite{Lenihan2022} for at least one of $U/t = 4$, $6$, $8$, excluded at that coupling; gray band: the $T/t=0.14$ isotherm.
        (b) Simple-cubic Brillouin zone: the path $\Gamma$--$X$--$M$--$R$--$\Gamma$ of Figs.~\ref{fig:mechanism}, \ref{fig:sigma_k} and~\ref{fig:partial_sums}; the nesting ridge $(\pi,\pi,q_z)$ with the spin-fluctuation peaks at $U/t=4$ on the $T/t=0.14$ isotherm (colored circles, at the grid maximum); and the noninteracting Fermi surface in the $k_z=0$ plane at $n=0.4$ and $0.9$.}
    \label{fig:map}
\end{figure}

The parameter set, Fig.~\ref{fig:map}(a), covers $n = 0.4$--$0.95$ at $\beta t = 3$, $4$, $5$, $7$, $10$, $20$ ($T/t = 0.33$--$0.05$), $48$ points in all.
The susceptibility points are the seven fillings $n = 0.4$--$0.9$ of the $T/t = 0.14$ isotherm and five points at other temperatures.

Our diagrammatic expansion (Sec.~\ref{sec:CDet}) starts from the paramagnetic state and describes the physical state only above the N\'eel temperature, which grows with $U$~\cite{Staudt2000,Kozik2013}.
The self-energy alone does not locate the transition, so we exclude the thirteen combinations of $(n,\beta,U)$, twelve of them at $n = 0.95$, that lie at or beyond the boundary computed with diagrammatic Monte Carlo in Ref.~\cite{Lenihan2022} [crosses in Fig.~\ref{fig:map}(a)].
Where we measure the susceptibility, it is consistent with this boundary at $U/t=4$; at $(n=0.9, T/t=0.14)$, where Ref.~\cite{Lenihan2022} places no order up to $U/t=8$, its peak at $U/t=6$ and $8$ cannot be distinguished from a divergence (Appendix~\ref{app:phase}).
The remaining $131$ combinations, $46$ of them at $U/t=4$, are called \emph{retained}.

\subsection{Connected-determinant expansion and resummation}
\label{sec:CDet}

We use the one-particle-irreducible CDet algorithm~\cite{Rossi2017,Moutenet2018,SimkovicKozik2019}, which samples the connected self-energy diagrams directly.
The simulations use the open-source FastFeynmanDiagrammatics codes for the self-energy and the susceptibilities~\cite{FFD}, which produced the two-dimensional results of Ref.~\cite{Simkovic2024Science} and which we extended to the simple-cubic lattice.
The expansion is built on the noninteracting propagator $G_0$ at the chemical potential $\mu_0$ that gives the target density.
The self-energy is expanded simultaneously in the interaction and in a chemical-potential shift $\alpha$,
\begin{equation}
    \Sigma(\mathbf{k},i\omega_n) = \sum_{j,l} \sigma_{j,l}(\mathbf{k},i\omega_n)\, U^j \alpha^l,
    \label{eq:double_expansion}
\end{equation}
and the same double expansion is measured for the spin and charge correlators.
The self-energy and the susceptibilities are sampled in separate simulations: the self-energy series reaches order $N_\mathrm{max} = 7$--$10$ in one simulation per $(n,\beta)$ point ($128$--$192$ threads for $48$ hours), and the susceptibility series $N_\mathrm{max} = 6$--$8$ at the twelve susceptibility points.
The coefficients are independent of $U$, so each simulation serves all three couplings; the fixed-density expansion~\cite{Simkovic2022,Simkovic2024Science} determines $\alpha(U)$ order by order so that the density stays at its target value.
The series starts at second order ($j=2$): the first-order Hartree term $Un/2$ is absorbed into the reference propagator $G_0$ and the counterterm into the chemical potential.
The reported $\Sigma$ therefore omits the constant $Un/2$, which affects neither its imaginary part nor its momentum dependence (Appendix~\ref{app:norm}).
Inserting $\alpha(U)$ gives the density-fixed series $\Sigma = \sum_j \sigma_j U^j$, with $\sigma_2 = \sigma_{2,0}$.
We extract the first eight fermionic and the first four bosonic Matsubara frequencies.

The series is truncated at the order set by the signal-to-noise ratio of its coefficients and resummed with Pad\'e, DLog-Pad\'e and Borel-Pad\'e estimators~\cite{Baker1996}; every reported number is the median of the converged estimators.
The systematic uncertainty of a resummed value combines the spread within and between the estimators with the shift on lowering the truncation order (the \emph{truncation term}), and its statistical uncertainty comes from resampling the Monte Carlo threads through the whole analysis (Appendix~\ref{app:resum}).
At $T/t \lesssim 0.2$ the resummation continues the series beyond its formal radius of convergence, set by a pairing singularity at negative $U$ that the computed orders do not resolve.
At $U/t=4$ a conformal map that opens the negative real axis reproduces the resummed scattering rate within half a systematic uncertainty, except at $T/t = 0.05$ and $n \geq 0.8$, where it lies up to two systematic uncertainties higher and the values are estimates (Appendix~\ref{app:singular}).
Quantities built from differences between momenta are resummed with all momenta at one \emph{common order}, at which the truncation error, nearly the same at every momentum, largely cancels (Appendix~\ref{app:estimators}).

\subsection{Single-particle observables}
\label{sec:quantities}

We characterize the self-energy at the lowest Matsubara frequencies, first through the scattering rate and the quasiparticle weight and then through the size of its momentum dependence.
At each momentum we extract the scattering rate
\begin{equation}
    \gamma(\mathbf{k}) = -\mathrm{Im}\,\Sigma(\mathbf{k}, i\omega_0),
    \label{eq:gamma}
\end{equation}
which on the Matsubara axis stands for the quasiparticle scattering rate at $\omega_0 = \pi T$.
In a Fermi liquid $\mathrm{Im}\,\Sigma(\mathbf k, i\omega_n) \approx -(Z^{-1}-1)\,\omega_n + A(\omega_n^2 - \pi^2T^2)$ at low positive frequency, and the second term vanishes at the first Matsubara frequency~\cite{Chubukov2012}, so that the quasiparticle weight
\begin{equation}
    Z_1(\mathbf{k}) = \bigl[1 + \gamma(\mathbf{k})/\pi T\bigr]^{-1}
    \label{eq:Z1}
\end{equation}
follows from the scattering rate alone.
We therefore also use the two-point estimator
\begin{equation}
    Z_\mathrm{2pt}^{-1}(\mathbf{k}) = 1 - \frac{\mathrm{Im}\,\Sigma(\mathbf{k},i\omega_1) - \mathrm{Im}\,\Sigma(\mathbf{k},i\omega_0)}{\omega_1 - \omega_0}
    \label{eq:Z}
\end{equation}
which measures the low-frequency slope independently of $\gamma$.
It is the estimator used for the momentum dependence of the quasiparticle weight, for the sign of the slope and for the comparison with DMFT.
The curvature of $\mathrm{Im}\,\Sigma$ between $\omega_0$ and $\omega_1 = 3\pi T$ biases it upward by a term of order $T$, so $Z_1$ is the better value of $Z$ itself (Sec.~\ref{sec:qp} and Appendix~\ref{app:z_freq}).
The local self-energy $\Sigma_\mathrm{loc}(i\omega_n) = N^{-1}\sum_{\mathbf k} w_{\mathbf k}\,\Sigma(\mathbf k, i\omega_n)$ is the zone average, with $w_\mathbf{k}$ the $O_h$ orbit multiplicity and $N = L^3$ the number of sites; $\gamma$, $Z_1$ and $Z_\mathrm{2pt}$ without a momentum argument refer to it, and $\bar\Sigma \equiv \Sigma_\mathrm{loc}(i\omega_0)$.

The momentum dependence of the self-energy is measured by its root-mean-square deviation from the local part at the lowest frequency,
\begin{equation}
    \Delta\Sigma_\mathrm{RMS} = \sqrt{\frac{1}{N}\sum_\mathbf{k} w_\mathbf{k}\, |\Sigma(\mathbf{k},i\omega_0) - \bar\Sigma|^2},
    \label{eq:nonlocality}
\end{equation}
quoted in units of $t$ together with its real and imaginary components.
We call $-\mathrm{Im}[\Sigma(\mathbf k,i\omega_0) - \bar\Sigma]$ the \emph{nonlocal scattering rate}; its size is the root-mean-square $\Delta\mathrm{Im}\Sigma_\mathrm{RMS}$, which we also quote relative to the local scattering rate, $\Delta\mathrm{Im}\Sigma_\mathrm{RMS}/|\mathrm{Im}\bar\Sigma|$.

Two references serve to interpret this momentum dependence.
The second-order term $U^2\sigma_{2,0}(\mathbf k,i\omega_0)$, the sunset diagram evaluated exactly with $G_0$ at $\mu_0$ on the same lattice, is the weak-coupling reference: the ratio of $\Delta\Sigma_\mathrm{RMS}$, or of one of its components, to that of the second-order term measures what lies beyond second order.
The band energy is the other: a self-energy that depends on $\mathbf k$ only through $\varepsilon_\mathbf{k}$, which we call isotropic, does not vary at fixed band energy, and the interacting Fermi surface is then itself a surface of constant band energy.
A real part linear in $\varepsilon_\mathbf{k}$, $c_1\eta_1(\mathbf k)$ with $\eta_1 = \sum_d \cos k_d = -\varepsilon_\mathbf{k}/2t$, renormalizes the nearest-neighbor hopping, $t_\mathrm{eff}/t - 1 = -c_1/2t$.
Table~\ref{tab:observables} collects these and the other observables, and Appendix~\ref{app:pt2} defines the fits and correlations used with them.

\begin{table}[tb]
    \centering
    \footnotesize
    \setlength{\tabcolsep}{3pt}
    \caption{Observables used in the text, with their definition and the place where they are defined in full.}
    \label{tab:observables}
    \begin{tabular}{lp{0.46\columnwidth}l}
        \toprule
        Symbol                                  & \raggedright Definition                                                                           & Where                                          \\
        \midrule
        $\gamma(\mathbf k)$, $\gamma$           & \raggedright scattering rate $-\mathrm{Im}\,\Sigma(\mathbf k,i\omega_0)$; zone average            & Eq.~\eqref{eq:gamma}                           \\
        $Z_1$, $Z_\mathrm{2pt}$                 & \raggedright first-Matsubara and two-point estimators of the quasiparticle weight                 & Eqs.~\eqref{eq:Z1}, \eqref{eq:Z}               \\
        $\Delta\Sigma_\mathrm{RMS}$             & \raggedright zone RMS of $\Sigma(\mathbf k,i\omega_0)-\bar\Sigma$ (Re, Im parts)                  & Eq.~\eqref{eq:nonlocality}                     \\
        $c_1$, $t_\mathrm{eff}$                 & \raggedright $\eta_1$ coefficient of $\mathrm{Re}\,\Sigma$; $t_\mathrm{eff}/t-1=-c_1/2t$          & Sec.~\ref{sec:quantities}                      \\
        $U_\mathrm{sp/ch}(\mathbf q,i\Omega_m)$ & \raggedright spin and charge vertices: random-phase form relative to $\chi_0$                     & Eq.~\eqref{eq:vertex_inv}                      \\
        $R(\mathbf q)$, $x^*$                   & \raggedright enhancement; zone-mean Stoner factor                                                 & Eqs.~\eqref{eq:enhancement}, \eqref{eq:stoner} \\
        $U_\mathrm{sp}(\mathbf q^*)$, $\bar U$  & \raggedright peak vertex; common peak vertex of the twelve points                                 & Secs.~\ref{sec:vertex_def}, \ref{sec:peak}     \\
        $D$, $\langle m_z^2\rangle$             & \raggedright double occupancy from the sum rules; $n-2D$                                          & Eq.~\eqref{eq:sumrules}                        \\
        $D_{\mathrm{Tr}\Sigma G}$               & \raggedright double occupancy from the Galitskii--Migdal trace of $\Sigma$                        & Appendix~\ref{app:docc}                        \\
        $U_\mathrm{eff}$                        & \raggedright TPSC spin sum rule solved with the sum-rule $D$                                      & Appendix~\ref{app:docc}                        \\
        $\xi$, $\xi_\mathrm{th}$                & \raggedright Ornstein--Zernike length across the ridge, Lorentzian width along it; thermal length & Sec.~\ref{sec:vertex_def}                      \\
        $\Sigma_\mathrm{cl}$                    & \raggedright $\Omega=0$ exchange of the measured $\chi_\mathrm{sp}$                               & Eq.~\eqref{eq:sigma_cl}                        \\
        \bottomrule
    \end{tabular}
\end{table}

\subsection{Susceptibilities, vertices and sum rules}
\label{sec:vertex_def}

The spin and charge susceptibilities are
\begin{align}
    \chi_\mathrm{sp/ch}(\mathbf{q}, i\Omega_m) ={} & \frac{1}{N}\sum_{i,j} e^{-i\mathbf{q}\cdot(\mathbf{r}_i-\mathbf{r}_j)} \nonumber                          \\
                                                   & \times \int_0^\beta\! d\tau\, e^{i\Omega_m\tau} \bigl\langle \hat{O}_i(\tau)\,\hat{O}_j(0)\bigr\rangle_c,
    \label{eq:chi_def}
\end{align}
with $\hat{O}_i = \hat{n}_{i\uparrow}\mp\hat{n}_{i\downarrow}$ for the spin (charge) channel, $\langle\cdots\rangle_c$ the connected average and $\Omega_m = 2\pi m T$; they are measured for $|m| \leq 3$, and a quantity without frequency argument is static, $\Omega_m = 0$.
Throughout, $\chi_0(\mathbf q, i\Omega_m)$ is the bare bubble of the same correlator, the noninteracting limit of Eq.~\eqref{eq:chi_def}, which in this normalization is twice the per-spin lattice Lindhard function; statically,
\begin{equation}
    \chi_0(\mathbf q) = -\frac{2}{N} \sum_\mathbf{k}
    \frac{f(\varepsilon_{\mathbf k + \mathbf q} - \mu_0) - f(\varepsilon_\mathbf k - \mu_0)}
    {\varepsilon_{\mathbf k + \mathbf q} - \varepsilon_\mathbf k},
    \label{eq:lindhard}
\end{equation}
evaluated on the same $L=20$ grid.
The measured susceptibilities are written in the random-phase form used by TPSC~\cite{Vilk1994,Vilk1997},
\begin{equation}
    \chi_\mathrm{sp} = \frac{\chi_0}{1 - U_\mathrm{sp}\chi_0/2}, \qquad
    \chi_\mathrm{ch} = \frac{\chi_0}{1 + U_\mathrm{ch}\chi_0/2},
    \label{eq:vertex_def}
\end{equation}
at each momentum and frequency, which defines the spin and charge vertices
\begin{align}
    U_\mathrm{sp}(\mathbf q, i\Omega_m) & = 2\Bigl[\frac{1}{\chi_0} - \frac{1}{\chi_\mathrm{sp}}\Bigr], \nonumber \\
    U_\mathrm{ch}(\mathbf q, i\Omega_m) & = -2\Bigl[\frac{1}{\chi_0} - \frac{1}{\chi_\mathrm{ch}}\Bigr].
    \label{eq:vertex_inv}
\end{align}
In this normalization the bare RPA is $U_\mathrm{sp} = U_\mathrm{ch} = U$, and TPSC vertices are directly comparable with Eq.~\eqref{eq:vertex_inv}.
Because it is defined relative to the bare bubble, the spin vertex is not the irreducible vertex of the Bethe--Salpeter equation, which D$\Gamma$A and related methods approximate, but a local-field factor that also absorbs the self-energy corrections to the bubble (Appendix~\ref{app:kb}).
We obtain the vertices by inverting the series of $\chi$ order by order into the series of the vertex and resumming that series (the \emph{vertex route}), which avoids the near-pole structure of $\chi$ close to an instability; resumming $\chi$ first (the \emph{direct route}) serves as a cross-check (Appendix~\ref{app:vertex_route}).

The enhancement at any momentum is fixed by the vertex at that momentum,
\begin{equation}
    R(\mathbf q) \equiv \frac{\chi_\mathrm{sp}(\mathbf q)}{\chi_0(\mathbf q)} = \frac{1}{1 - U_\mathrm{sp}(\mathbf q)\chi_0(\mathbf q)/2}.
    \label{eq:enhancement}
\end{equation}
For a given bubble the height of the spin-fluctuation peak therefore depends on the vertex only through the peak vertex $U_\mathrm{sp}(\mathbf q^*)$, where the peak momentum $\mathbf q^*$ is the grid momentum at which $\chi_\mathrm{sp}$ is largest on the nesting ridge $(\pi,\pi,q_z)$.
A Lorentzian fit to the ridge values locates the peak position $q_z^\mathrm{fit}$ between grid points (Appendix~\ref{app:corr}).
Near an instability the Stoner denominator $1 - U_\mathrm{sp}\chi_0/2$ is small, and it amplifies small differences of this vertex.
To measure how much the peak depends on the vertex there rather than on its average, we compare $R(\mathbf q^*)$ with $1/(1-x^*)$, where
\begin{equation}
    x^* = \langle U_\mathrm{sp}\rangle_\mathbf{q}\,\chi_0(\mathbf q^*)/2
    \label{eq:stoner}
\end{equation}
is the Stoner factor of the zone-mean vertex, with $\langle\cdot\rangle_\mathbf q$ the $O_h$-weighted Brillouin-zone average.
The spin correlation length $\xi$ is the Ornstein--Zernike length from the curvature of $1/\chi_\mathrm{sp}$ around $\mathbf q^*$, $1/\chi_\mathrm{sp} \approx a + b\sum_d 4\sin^2(\delta q_d/2)$ and $\xi = \sqrt{b/a}$, taken across the ridge; along the ridge the width of the Lorentzian fit that locates the peak (Appendix~\ref{app:corr}) plays its role, and for $\xi \lesssim 1$ either measures the sharpness of the peak rather than a decay length.
It is compared with the thermal length $\xi_\mathrm{th} = \langle v_F\rangle/\pi T$, with $\langle v_F\rangle$ the Fermi-surface average of the band velocity, because in two dimensions spin fluctuations in the renormalized-classical regime produce hot spots and a pseudogap once $\xi \gg \xi_\mathrm{th}$~\cite{Vilk1997}.

The local sum rules connect the susceptibilities to the double occupancy $D = \langle n_{i\uparrow}n_{i\downarrow}\rangle$:
\begin{align}
    \frac{T}{N}\sum_{\mathbf q, m}\chi_\mathrm{sp}(\mathbf q,i\Omega_m) & = n - 2D, \nonumber \\
    \frac{T}{N}\sum_{\mathbf q, m}\chi_\mathrm{ch}(\mathbf q,i\Omega_m) & = n + 2D - n^2,
    \label{eq:sumrules}
\end{align}
where $n - 2D = \langle m_z^2\rangle$ is the local moment, $\langle m_z^2\rangle_0 = n - n^2/2$ without interaction.
TPSC replaces the two vertices by constants $U^\mathrm{TPSC}_\mathrm{sp}$, $U^\mathrm{TPSC}_\mathrm{ch}$ that satisfy Eq.~\eqref{eq:sumrules} with the ansatz $D = (U_\mathrm{sp}/U)(n/2)^2$~\cite{Vilk1997,Dare1996,Arita2000}, evaluated here on the same grid and at the same $\mu_0$.

\subsection{Precision and checks}
\label{sec:precision}

At $U/t=4$ the resummed series are well controlled: the median uncertainty of the spin vertex is $2.5\%$ over the zone, the local scattering rate has a median truncation term of $0.7\%$ for $n \leq 0.8$, and the nonlocal scattering rate, a small difference between momenta, carries a median systematic uncertainty of $12\%$ (Appendices~\ref{app:sys} and~\ref{app:vertex_route}).
The truncation error of the nonlocal scattering rate is bounded at all but two of the points with $n = 0.7$--$0.9$ and $T/t \geq 0.14$, which carry the main statements on the nonlocal self-energy in Sec.~\ref{sec:sigma}; at the $26$ points with $n \leq 0.6$, where the nonlocal part is small, or with $T/t \leq 0.10$, it is bounded at only two, and the values quoted there are estimates.
At $U/t = 6$ and $8$ the median truncation term of the local scattering rate grows to $2\%$ and $6\%$ for $n \leq 0.8$, and the median systematic uncertainty of the nonlocal scattering rate to $52\%$ and about $100\%$, so the nonlocal self-energy shows trends only.
There the spin vertex enters only through its zone mean, and the spin-fluctuation peak is not analyzed.
At nine of the twenty-four combinations of the susceptibility points with these couplings its height cannot be distinguished from a divergence within one standard deviation.
These include, at both couplings, the points with $n \geq 0.85$ at $T/t=0.14$ and the point at $T/t=0.05$, so the peak cannot be followed to high filling on the isotherm or to low temperature.

Runs at $L=24$ at four points change $Z_\mathrm{2pt}$ by at most $0.3\%$ and $\gamma$ by at most $1\%$ for $U/t \leq 6$, and the momentum dependence of the scattering rate by less than its uncertainty; a susceptibility run at $L=24$ leaves the static vertex unchanged within its uncertainty.
For $T/t \geq 0.10$ the exact size dependence of the bare bubble also keeps the finite-size effect on the peak susceptibility below its uncertainty, whereas at $T/t=0.05$ the finite-size effects reach a few percent in the scattering rate and $10\%$ in the peak susceptibility (Appendix~\ref{app:finite_size}).

Three further checks compare the results with independent determinations.
The computed coefficients reproduce the exact low-order ones on the same lattice: the second- and third-order self-energy, the lowest-order coefficients of the susceptibility and vertex series, and the second-order coefficient of the spin vertex (Appendices~\ref{app:norm}, \ref{app:vertex_route}, \ref{app:kb} and~\ref{app:h3}).
The double occupancies from the resummed self-energy and, through the sum rules, from the resummed susceptibilities, which come from separate simulations and are resummed independently, agree within $0.4$ combined standard deviations at the nine points where both are available (Sec.~\ref{sec:localmoment}).
Two independent runs at $(n=0.7, \beta t=7)$ reproduce $Z_\mathrm{2pt}$ and $\gamma$ to within $0.04\%$ and $0.3\%$ at $U/t=4$ and to within $0.35\%$ and $1.8\%$ at $U/t=8$ (Appendix~\ref{app:stat}).

\section{The spin vertex and the spin-fluctuation peak}
\label{sec:vertex}

This section shows that the spin fluctuations are strong yet short ranged, and that their strength is controlled by a screened, nearly uniform spin vertex whose dip on the nesting ridge sets the height of the peak.

\begin{figure}[!t]
    \centering
    \includegraphics[width=\columnwidth]{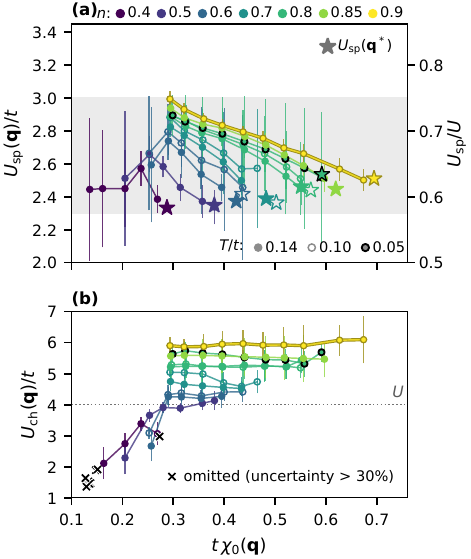}
    \caption{The static spin vertex is screened to about two thirds of $U$ and nearly uniform over the zone, lowest where the bare bubble is largest; the charge vertex grows with filling.
        Static vertices at $U/t=4$ against the bare bubble $\chi_0(\mathbf q)$ at the same momentum for the eleven susceptibility points with $T/t \leq 0.14$, as means over bins of width $0.04/t$ in $\chi_0$; bars: the mean single-momentum uncertainty in the bin.
        (a) Spin vertex $U_\mathrm{sp}(\mathbf q)$, on the right axis in units of $U$; stars: the peak vertex $U_\mathrm{sp}(\mathbf q^*)$; gray band: $2.3t \leq U_\mathrm{sp} \leq 3.0t$.
        (b) Charge vertex $U_\mathrm{ch}(\mathbf q)$; dotted: the bare $U$; crosses: the nine cells omitted because their relative uncertainty exceeds $30\%$ (Appendix~\ref{app:vertex_route}; none in the spin channel).
        Color encodes filling; fill encodes temperature: filled $T/t=0.14$, open $0.10$, black-edged $0.05$.}
    \label{fig:vertex_q}
\end{figure}

\subsection{Screening and a nearly uniform spin vertex}
\label{sec:vertex_sp}

At $U/t=4$ the static spin vertex is strongly screened and nearly momentum independent.
In all but $12$ of the $3431$ momentum--point cells (Appendix~\ref{app:vertex_route}) it lies between $2.3t$ and $3.0t$, while the susceptibility it parametrizes varies by more than an order of magnitude [Fig.~\ref{fig:vertex_q}(a)].
Its zone mean, about $0.68\,U$, rises only from $2.55t$ at $n=0.4$ to $2.87t$ at $n=0.9$ (Table~\ref{tab:vertex}).

The screening is compatible with the Kanamori--Brueckner mechanism~\cite{Kanamori1963,Chen1991,Vilk1994}.
At second order in $U$, where it can be resolved exactly at the seven $T/t=0.14$ points, particle-particle scattering provides $65$--$78\%$ of the reduction of the zone mean; the particle-hole vertex corrections cancel identically in the spin channel, and the rest comes from self-energy corrections to the bubble (Appendix~\ref{app:kb}).
With increasing coupling the zone mean falls to $0.50\,U$ at $U/t=8$.
The Kanamori form $U/(1+U/U_\mathrm{K})$ describes this within $2\%$ with $U_\mathrm{K} = 8.2t$, close to the $7.9t$ of the $T$-matrix at the band bottom, but it does not follow the filling dependence of the vertex (Appendix~\ref{app:kb}).

\begin{table*}[tb]
\centering
\footnotesize
\setlength{\tabcolsep}{4pt}
\caption{Renormalized vertices at $U/t=4$ at the twelve susceptibility points (Sec.~\ref{sec:vertex}; vertices in units of $t$, $B$ in units of $t^2$).
$\langle U_\mathrm{sp}\rangle$, $\langle U_\mathrm{ch}\rangle$: Brillouin-zone averages of $U_\mathrm{sp}(\mathbf q)$ and $U_\mathrm{ch}(\mathbf q)$, Eq.~\eqref{eq:vertex_inv} (the $286$ irreducible momenta weighted by their $O_h$ multiplicity), followed in brackets by the $5$--$95\%$ range over the zone; the median uncertainty of $\langle U_\mathrm{sp}\rangle$ over the twelve points is $3.0\%$ (Appendix~\ref{app:vertex_route}).
$B$: slope of the linear fit $U_\mathrm{sp}(\mathbf q)=U_0-B\,\chi_0(\mathbf q)$ over the zone, with the least-squares [scaled by $(\chi^2/\mathrm{dof})^{1/2}$ where this exceeds one], the thread-bootstrap and the truncation [$|B(N-1)-B(N)|$, Appendix~\ref{app:resum}] uncertainties combined in quadrature.
$U_\mathrm{sp}(\mathbf q^*)$: spin vertex at the grid maximum $\mathbf q^*$ of $\chi_\mathrm{sp}$ on the $(\pi,\pi,q_z)$ ridge (Sec.~\ref{sec:vertex_def}).
TPSC: parameter-free vertices of Ref.~\cite{Vilk1997} evaluated at the same $(n,T)$.
Uncertainties in parentheses combine the resummation spread, the truncation systematic [for $U_\mathrm{sp}(\mathbf q^*)$ the term of Eq.~\eqref{eq:trunc_rule}] and the statistical error in quadrature, in units of the last digit (Appendix~\ref{app:resum}).}
\label{tab:vertex}
\begin{tabular}{ccc ccc c cc}
\toprule
$n$ & $\beta t$ & $T/t$ & $\langle U_\mathrm{sp}\rangle\,[5\%,95\%]$ & $B$ & $U_\mathrm{sp}(\mathbf q^*)$ & $U_\mathrm{sp}^\mathrm{TPSC}$ & $\langle U_\mathrm{ch}\rangle\,[5\%,95\%]$ & $U_\mathrm{ch}^\mathrm{TPSC}$ \\
\midrule
0.4 & 7 & 0.14 & $2.55\,[2.38,2.67]$ & 1.4(6) & 2.33(9) & 2.13 & $3.28\,[2.58,3.61]$ & 3.15 \\
0.5 & 7 & 0.14 & $2.57\,[2.38,2.71]$ & 3.87(34) & 2.35(11) & 2.17 & $3.85\,[2.98,4.15]$ & 3.64 \\
0.6 & 7 & 0.14 & $2.66\,[2.46,2.77]$ & 2.90(31) & 2.37(12) & 2.24 & $4.24\,[3.93,4.51]$ & 4.25 \\
0.6 & 10 & 0.10 & $2.72\,[2.56,2.83]$ & 3.2(12) & 2.4(5) & 2.25 & $4.34\,[4.04,4.54]$ & 4.23 \\
0.7 & 7 & 0.14 & $2.72\,[2.55,2.82]$ & 2.2(4) & 2.39(12) & 2.32 & $4.63\,[4.48,4.85]$ & 5.02 \\
0.7 & 10 & 0.10 & $2.76\,[2.61,2.85]$ & 2.2(7) & 2.36(14) & 2.33 & $4.97\,[4.60,5.18]$ & 4.99 \\
0.8 & 7 & 0.14 & $2.78\,[2.66,2.88]$ & 1.30(10) & 2.46(9) & 2.41 & $5.24\,[5.17,5.34]$ & 6.03 \\
0.8 & 10 & 0.10 & $2.80\,[2.67,2.90]$ & 1.6(7) & 2.44(14) & 2.42 & $5.26\,[5.17,5.37]$ & 5.98 \\
0.8 & 20 & 0.05 & $2.81\,[2.70,2.91]$ & 1.1(7) & 2.5(4) & 2.43 & $5.66\,[5.28,5.88]$ & 5.92 \\
0.85 & 7 & 0.14 & $2.83\,[2.73,2.92]$ & 1.06(23) & 2.45(10) & 2.46 & $5.58\,[5.50,5.70]$ & 6.64 \\
0.9 & 3 & 0.33 & $2.80\,[2.68,2.90]$ & 1.27(5) & 2.430(6) & 2.44 & $6.16\,[6.07,6.25]$ & 7.67 \\
0.9 & 7 & 0.14 & $2.87\,[2.75,2.98]$ & 1.17(11) & 2.51(10) & 2.51 & $5.90\,[5.76,6.04]$ & 7.35 \\
\bottomrule
\end{tabular}
\end{table*}

The spread of the vertex is small but systematic.
The vertex is smallest where the bare bubble is largest: a straight line $U_\mathrm{sp}(\mathbf q) = U_0 - B\,\chi_0(\mathbf q)$ (slopes $B$ in Table~\ref{tab:vertex}) describes it within the precision of a single momentum (except at $(n=0.9, T/t=0.33)$, where the vertex is known to about $0.4\%$ and departs from the line by about $1.5\%$ root mean square), so that
\begin{equation}
    \frac{1}{R(\mathbf q)} = 1 - \frac{U_0\chi_0(\mathbf q)}{2} + \frac{B\chi_0^2(\mathbf q)}{2}.
    \label{eq:linearlaw}
\end{equation}
The measured slopes keep $B\chi_0^2$ below about $0.6$, far from the value $2$ above which $\chi_\mathrm{sp}$ would decrease with $\chi_0$, so $\chi_\mathrm{sp}$ increases monotonically with $\chi_0$.
Since the largest bubble lies on the nesting ridge $(\pi,\pi,q_z)$, the vertex dips there: the peak vertex is $U_\mathrm{sp}(\mathbf q^*) = 2.33$--$2.51t$, about $0.6\,U$, and the zone mean lies $9$--$17\%$ higher.
Referenced to a bubble built from the CDet self-energy, the dip survives and deepens at $n = 0.7$--$0.8$, where at second order it is carried by particle-particle scattering; at $n = 0.4$ it disappears, so at low filling the dip reflects the self-energy corrections to the bubble (Appendix~\ref{app:kb}).

The charge vertex behaves differently.
At $U/t=4$ its zone mean rises from $3.3t$ at $n=0.4$ to $5.9t$ at $n=0.9$, exceeding the bare $U$ near half filling [Fig.~\ref{fig:vertex_q}(b), Table~\ref{tab:vertex}], as TPSC and quantum Monte Carlo found in two dimensions~\cite{Vilk1994}, while the maximum of $\chi_\mathrm{ch}$ over the zone stays at $0.20$--$0.23$.
Repulsion suppresses charge fluctuations to a nearly filling-independent level, and Eq.~\eqref{eq:vertex_inv} converts a growing $\chi_0$ at fixed $\chi_\mathrm{ch}$ into a growing vertex; in the static response the collective enhancement appears in the spin channel only.

\subsection{Position, width and height of the peak}
\label{sec:peak}

\begin{figure*}[t]
    \centering
    \includegraphics[width=\textwidth]{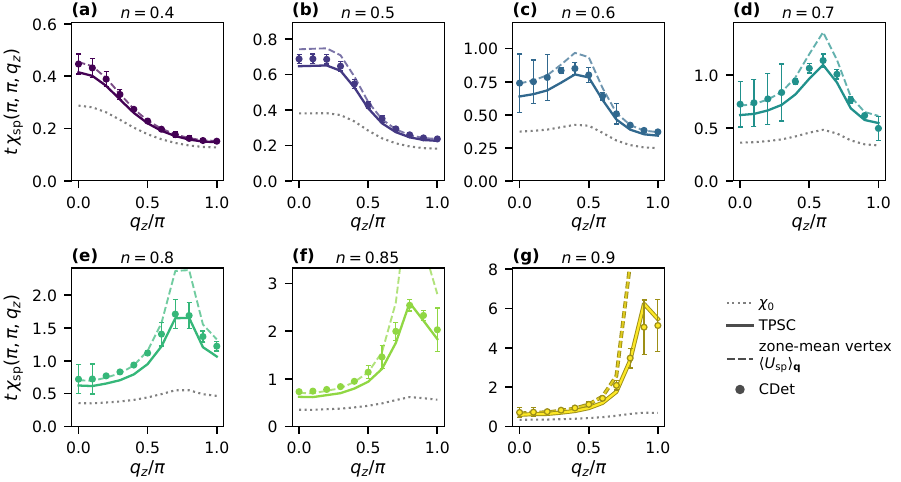}
    \caption{With increasing filling the spin-fluctuation peak grows and moves along the nesting ridge; TPSC follows it, whereas a constant vertex equal to the zone mean overshoots it increasingly toward half filling.
        Static spin susceptibility along $\mathbf q = (\pi,\pi,q_z)$ at $U/t=4$ for the seven fillings of the $T/t=0.14$ isotherm.
        Circles: CDet, direct route (bars: spread of the resummation estimators and truncation term); dotted gray: the bare bubble $\chi_0$, Eq.~\eqref{eq:lindhard}; solid: TPSC, $\chi_0/(1 - U_\mathrm{sp}^\mathrm{TPSC}\chi_0/2)$; dashed: the same form with the zone-mean vertex $\langle U_\mathrm{sp}\rangle_\mathbf q$ (blanked where it diverges).}
    \label{fig:ridge}
\end{figure*}

\begin{figure}[!t]
    \centering
    \includegraphics[width=\columnwidth]{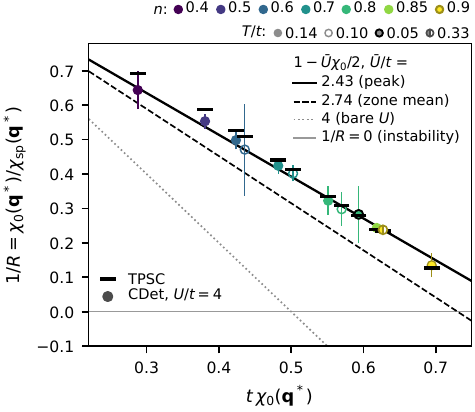}
    \caption{The twelve peak heights fall on one line set by a common peak vertex, which lies below the zone-mean vertex.
        Inverse enhancement $1/R = \chi_0(\mathbf q^*)/\chi_\mathrm{sp}(\mathbf q^*)$ against the bare bubble at the peak for the twelve points at $U/t=4$ (circles).
        Lines: $1 - \bar U\chi_0/2$ with the common peak vertex $\bar U = 2.43t$ (solid, direct route), with the average zone-mean vertex, $2.74t$ (dashed), and with the bare $U$ (dotted); gray: $1/R = 0$, the Stoner instability.
        Thick bars: the TPSC prediction at each point.
        Color encodes filling (key above the panel); fill encodes temperature: filled $T/t=0.14$, open $0.10$, black-edged $0.05$, half-filled $0.33$.}
    \label{fig:peak}
\end{figure}

\begin{table}[tb]
\centering
\footnotesize
\setlength{\tabcolsep}{1.6pt}
\caption{Position and height of the spin-fluctuation peak at $U/t=4$ (Sec.~\ref{sec:peak}).
$q_z^\mathrm{fit}$: position of the maximum of $\chi_\mathrm{sp}$ along $(\pi,\pi,q_z)$ in units of $\pi$, from a Lorentzian fit (Appendix~\ref{app:corr}); $q_{z,0}^\mathrm{fit}$: the same for $\chi_0$.
$\chi_0(\mathbf q^*)$: bare bubble, in units of $1/t$, at the grid maximum $\mathbf q^*$ of $\chi_\mathrm{sp}$ on the ridge, which at $U/t=4$ is also the grid maximum of $\chi_0$; $x^*=\langle U_\mathrm{sp}\rangle\chi_0(\mathbf q^*)/2$: Stoner factor of the zone-mean vertex, to three decimals where $x^* \geq 0.95$.
$R$: measured enhancement $\chi_\mathrm{sp}(\mathbf q^*)/\chi_0(\mathbf q^*)$ (resummation and truncation uncertainty; $^{\mathrm g}$: $-7.3\%$ at $(n=0.8,\beta t=20)$ with the $L=100$ bubble (Appendix~\ref{app:finite_size}), not included); $R_\mathrm{TPSC}$: TPSC prediction; $1/(1-x^*)$: prediction of a constant vertex equal to the zone mean, from the central $x^*$; $x^*=0.996(11)$ at $(n=0.9,\beta t=7)$ (zone-mean uncertainty, Appendix~\ref{app:vertex_route}): that prediction is divergent within errors.}
\label{tab:peak}
\begin{tabular}{cc cc cc ccc}
\toprule
$n$ & $\beta t$ & $q_z^\mathrm{fit}$ & $q_{z,0}^\mathrm{fit}$ & $\chi_0(\mathbf q^*)$ & $x^*$ & $R$ & $R_\mathrm{TPSC}$ & $1/(1-x^*)$ \\
\midrule
0.4 & 7 & 0.08(9) & 0.04 & 0.288 & 0.37 & 1.55(13) & 1.44 & 1.58 \\
0.5 & 7 & 0.12(6) & 0.14 & 0.381 & 0.49 & 1.81(7) & 1.70 & 1.96 \\
0.6 & 7 & 0.29(7) & 0.31 & 0.424 & 0.56 & 2.01(11) & 1.90 & 2.29 \\
0.6 & 10 & 0.27(8) & 0.33 & 0.436 & 0.59 & 2.1(6) & 1.96 & 2.45 \\
0.7 & 7 & 0.56(7) & 0.56 & 0.482 & 0.66 & 2.36(13) & 2.27 & 2.91 \\
0.7 & 10 & 0.54(10) & 0.57 & 0.503 & 0.69 & 2.49(14) & 2.42 & 3.27 \\
0.8 & 7 & 0.76(8) & 0.76 & 0.551 & 0.77 & 3.1(4) & 2.98 & 4.29 \\
0.8 & 10 & 0.73(6) & 0.75 & 0.570 & 0.80 & 3.4(5) & 3.23 & 4.97 \\
0.8 & 20 & 0.73(15) & 0.74 & 0.593 & 0.83 & 3.5(10)$^{\mathrm g}$ & 3.57 & 6.00 \\
0.85 & 7 & 0.83(5) & 0.85 & 0.618 & 0.88 & 4.11(19) & 4.20 & 8.07 \\
0.9 & 3 & 0.95(6) & 0.97 & 0.627 & 0.88 & 4.20(23) & 4.26 & 8.10 \\
0.9 & 7 & 0.94(8) & 0.93 & 0.694 & 0.996 & 7.4(19) & 7.88 & 229 \\
\bottomrule
\end{tabular}
\end{table}

At all twelve points the maximum of $\chi_\mathrm{sp}$ lies on the nesting ridge at the maximum of the bare bubble (Fig.~\ref{fig:ridge}, Table~\ref{tab:peak}), as the monotonic dependence of $\chi_\mathrm{sp}$ on $\chi_0$ requires.
Along the $T/t=0.14$ isotherm the peak moves from $q_z^\mathrm{fit} = 0.08(9)\pi$ at $n=0.4$ to $0.94(8)\pi$ at $n=0.9$ [Table~\ref{tab:peak}; grid maxima in Fig.~\ref{fig:map}(b)], within $0.04\pi$ of the bubble maximum.
This is the path predicted by D$\Gamma$A on the instability line~\cite{Schafer2017}, and DMFT near the magnetic quantum critical point also places the maximum on this ridge~\cite{Fus2024}.
At $T/t=0.05$ the static peak at $n=0.8$ stays near $q_z^\mathrm{fit} = 0.73(15)\pi$, as at higher temperature, consistent with the incommensurate correlations reported in the equal-time structure factor~\cite{Lenihan2022}.

The peak is broad, so the fluctuations are short ranged.
Across the ridge $\xi = 0.4$--$1.4$ lattice spacings at the twelve points, and along it at most about two, well below the thermal length $\xi_\mathrm{th}$, which is $4.9$--$5.8$ lattice spacings at $T/t=0.14$; the ratio $\xi/\xi_\mathrm{th}$ stays below one half in every direction, and across the ridge at most $0.24$ for $T/t \leq 0.14$ (Appendix~\ref{app:corr}).

The height of the peak is set by a peak vertex that hardly changes.
The measured enhancement $R(\mathbf q^*)$ rises from $1.55(13)$ at $n=0.4$ to $7.4(19)$ at $(n=0.9, T/t=0.14)$, while the bare bubble at the peak grows only by a factor of $2.4$ (Table~\ref{tab:peak}).
Plotted as $1/R$ against $\chi_0(\mathbf q^*)$ (Fig.~\ref{fig:peak}), the twelve peaks fall on one line $1 - \bar U\chi_0/2$ with the common peak vertex $\bar U = 2.43(7)t$.
Lines with different $\bar U$ separate in proportion to $\chi_0$, so the points near half filling fix $\bar U$ most tightly, while those with $n \leq 0.6$, where $R \approx 1.5$--$2$, test that the same value holds at small $\chi_0$.
All twelve scatter about the line by less than their individual uncertainties, typically $5\%$ (Appendix~\ref{app:zone_err}).

Because the height is set by the vertex on the ridge, a theory with a single vertex must match that value rather than the zone average.
The Stoner denominator amplifies the difference from the zone mean: once $x^*$ approaches $0.9$, $1/(1-x^*)$ is about twice the measured enhancement, and at $(n=0.9, T/t=0.14)$ it diverges within errors (dashed lines in Fig.~\ref{fig:ridge}).
Bare-$U$ RPA exceeds its Stoner limit at seven of the twelve points.

TPSC meets this requirement.
Its parameter-free vertex lies close to the peak vertex: on the isotherm their ratio rises from $0.91$ at $n=0.4$ to $1.00$ at $n=0.9$, while the TPSC vertex stays $12$--$17\%$ below the zone mean.
TPSC therefore reproduces the measured enhancement within $1.6$ standard deviations at all twelve points (thick bars in Fig.~\ref{fig:peak}), and the agreement grows with the expansion order (Appendix~\ref{app:zone_err}).
At low filling this agreement says little, since a relative error of the vertex enters the enhancement multiplied by $R - 1 \lesssim 1$.
For $n \geq 0.7$ it constrains the vertex closely, except at $(n=0.9, T/t=0.14)$ and $(0.8, 0.05)$, where the measured enhancement is itself uncertain by more than a quarter.
Section~\ref{sec:localmoment} shows why the TPSC vertex lands near the peak vertex.

\subsection{Frequency dependence, local moment and the sum rule}
\label{sec:localmoment}

\begin{figure*}[t]
    \centering
    \includegraphics[width=\textwidth]{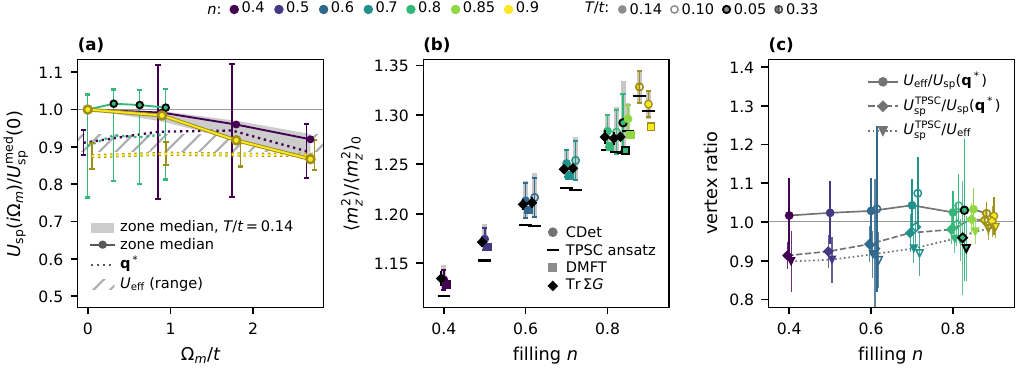}
    \caption{The spin vertex becomes flat in $\mathbf q$ at finite frequency, the local moment is enhanced, and the sum-rule vertex lands near the static peak vertex ($U/t=4$).
        (a) $U_\mathrm{sp}(i\Omega_m)$ divided by the static zone median: zone median (solid) and value at $\mathbf q^*$ (dotted; bars: its per-frequency uncertainty, offset in $\Omega_m$ for clarity) at $(n, T/t) = (0.4, 0.14)$, $(0.9, 0.14)$ and $(0.8, 0.05)$; gray band: range of the zone median on the $T/t=0.14$ isotherm; hatched band: range of $U_\mathrm{eff}$ over the static zone median at the twelve points.
        (b) $\langle m_z^2\rangle/\langle m_z^2\rangle_0$ from the sum-rule $D$ (circles; light bars include the tail model, Appendix~\ref{app:docc}), the TPSC ansatz (black bars), single-site DMFT (squares) and the Galitskii--Migdal trace (diamonds, $n \leq 0.8$).
        (c) $U_\mathrm{eff}/U_\mathrm{sp}(\mathbf q^*)$ (circles), $U^\mathrm{TPSC}_\mathrm{sp}/U_\mathrm{sp}(\mathbf q^*)$ (diamonds) and $U^\mathrm{TPSC}_\mathrm{sp}/U_\mathrm{eff}$ (down-triangles), with bars from the uncertainties of $U_\mathrm{sp}(\mathbf q^*)$ and $U_\mathrm{eff}$ ($U^\mathrm{TPSC}_\mathrm{sp}$ carries none).
        Color: filling; fill: temperature as in Fig.~\ref{fig:peak}. Values in Table~\ref{tab:docc}.}
    \label{fig:localmoment}
\end{figure*}

The dip of the spin vertex on the ridge is a static effect.
Between $\Omega_0$ and $\Omega_3 = 2.7t$ the zone median of $U_\mathrm{sp}(\mathbf q, i\Omega_m)$, which unlike the mean needs no precision cut on the noisier momenta at finite frequency, falls by $7$--$13\%$ at the $T/t=0.14$ points, while the peak vertex stays constant within its per-frequency uncertainty of $2$--$19\%$ [bars in Fig.~\ref{fig:localmoment}(a)].
Their ratio rises from $0.86$--$0.91$ to $0.97$--$1.05$: at finite frequency the vertex is flat in $\mathbf q$.
Part of the decrease with frequency reflects the bare-bubble reference, which ignores the reduced quasiparticle weight and kinetic energy (Appendix~\ref{app:docc}).

The local moment is enhanced.
The interaction suppresses the double occupancy $D$ to $47$--$62\%$ of its noninteracting value $(n/2)^2$, so $\langle m_z^2\rangle/\langle m_z^2\rangle_0$ rises from $1.13$ at $n=0.4$ to $1.31$ at $n=0.9$ on the isotherm ($1.33$ at $T/t=0.33$) [Fig.~\ref{fig:localmoment}(b)].
This $D$ follows from the sum rules, Eq.~\eqref{eq:sumrules}, with the measured susceptibilities at $|m| \leq 3$ and a modeled high-frequency tail whose uncertainty is quoted separately (Appendix~\ref{app:docc}).
The Galitskii--Migdal trace of the resummed self-energy, an independent route, confirms it within $0.4$ combined standard deviations at the nine points with $n \leq 0.8$, and at $T/t \geq 0.10$ its uncertainty is $2$--$8$ times smaller (Table~\ref{tab:docc}).
Single-site DMFT places $D$ a few percent above the Galitskii--Migdal value, within $1.6$ standard deviations at $T/t=0.14$ and $n \leq 0.8$.

\begin{table*}[tb]
\centering
\footnotesize
\setlength{\tabcolsep}{5pt}
\caption{Double occupancy $D=\langle n_{i\uparrow}n_{i\downarrow}\rangle$ from the local spin and charge sum rules and the sum-rule spin vertex at $U/t=4$, from the measured $\chi_\mathrm{sp/ch}(\mathbf q,i\Omega_m)$ at $|m|\leq 3$ with a per-momentum RPA-form tail beyond (Appendix~\ref{app:docc}). $D$: total uncertainty (statistical, resummation, route and tail model), and in braces its tail-model part, in units of the last digit; $D/(n/2)^2$ shares its relative uncertainty, and $D_\mathrm{ans}/D_{\mathrm{Tr}\Sigma G}$ and $D_\mathrm{DMFT}/D_{\mathrm{Tr}\Sigma G}$ that of $D_{\mathrm{Tr}\Sigma G}$. $D_{\mathrm{Tr}\Sigma G}$: Galitskii--Migdal trace of the resummed self-energy (Appendix~\ref{app:docc}). $D_\mathrm{ans}=(U_\mathrm{sp}^\mathrm{TPSC}/U)(n/2)^2$: TPSC ansatz; $D_\mathrm{DMFT}$: double occupancy of the single-site DMFT impurity (Appendix~\ref{app:dmft}); $U_\mathrm{eff}$: TPSC spin sum rule solved with the sum-rule $D$; --: not evaluated.}
\label{tab:docc}
\begin{tabular}{cc cc c ccc cc}
\toprule
$n$ & $T/t$ & $D$ & $D_{\mathrm{Tr}\Sigma G}$ & $\dfrac{D}{(n/2)^2}$ & $\dfrac{D_\mathrm{ans}}{D_{\mathrm{Tr}\Sigma G}}$ & $\dfrac{D_\mathrm{DMFT}}{D_{\mathrm{Tr}\Sigma G}}$ & $\dfrac{U_\mathrm{eff}}{U_\mathrm{sp}(\mathbf q^*)}$ & $\dfrac{U_\mathrm{sp}^\mathrm{TPSC}}{U_\mathrm{eff}}$ \\
\midrule
0.4 & 0.14 & 0.0187(23)\{17\} & 0.0185(12) & 0.467 & 1.152 & 1.054 & 1.02(10) & 0.90(8) \\
0.5 & 0.14 & 0.0298(28)\{19\} & 0.0304(10) & 0.477 & 1.116 & 1.031 & 1.02(8) & 0.90(6) \\
0.6 & 0.14 & 0.0453(42)\{20\} & 0.0460(11) & 0.503 & 1.094 & 1.026 & 1.03(9) & 0.92(6) \\
0.6 & 0.10 & 0.0446(52)\{31\} & 0.0457(12) & 0.495 & 1.109 & -- & 1.03(22) & 0.90(7) \\
0.7 & 0.14 & 0.0654(34)\{15\} & 0.0668(9) & 0.534 & 1.065 & 1.022 & 1.04(7) & 0.93(4) \\
0.7 & 0.10 & 0.0647(54)\{29\} & 0.0666(7) & 0.528 & 1.074 & -- & 1.07(9) & 0.92(6) \\
0.8 & 0.14 & 0.0920(42)\{15\} & 0.0934(16) & 0.575 & 1.034 & 1.025 & 1.02(6) & 0.96(4) \\
0.8 & 0.10 & 0.0922(62)\{31\} & 0.0937(15) & 0.576 & 1.035 & -- & 1.04(8) & 0.95(5) \\
0.8 & 0.05 & 0.0899(102)\{75\} & 0.0933(8) & 0.562 & 1.043 & 1.036 & 1.03(18) & 0.93(8) \\
0.85 & 0.14 & 0.1083(39)\{17\} & -- & 0.600 & -- & -- & 1.03(5) & 0.98(3) \\
0.9 & 0.33 & 0.1213(41)\{5\} & -- & 0.599 & -- & -- & 1.02(3) & 0.98(3) \\
0.9 & 0.14 & 0.1256(38)\{20\} & -- & 0.620 & -- & -- & 1.02(5) & 0.99(3) \\
\bottomrule
\end{tabular}
\end{table*}

This $D$ shows why TPSC succeeds at the peak [Fig.~\ref{fig:localmoment}(c)].
Fed the measured $D$ instead of its ansatz, the TPSC spin sum rule, whose sensitivity to the vertex lies mostly at the computed frequencies $|m| \leq 3$, returns a constant $U_\mathrm{eff} = 1.03(4)\,U_\mathrm{sp}(\mathbf q^*)$ with no filling trend.
The sum rule averages the vertex over all momenta and frequencies and is dominated by those away from the peak (Appendix~\ref{app:docc}), so one would expect its constant near the static zone mean, which exceeds the ridge value by $9$--$17\%$.
Two reductions of similar size bring the constant down.
First, the vertex falls with frequency: weighted by the sensitivity of the sum rule to it, the vertex at $|m| \leq 3$ lies $2$--$5\%$ below the static zone mean at the $T/t=0.14$ points.
Second, the high-frequency tail of the measured $\chi_\mathrm{sp}$, whose $1/\Omega^2$ amplitude is proportional to the kinetic energy, is smaller than that of the bare bubble.
A constant vertex on the bare bubble compensates for the larger bubble tail by being another $4$--$7\%$ smaller (Appendix~\ref{app:docc}).
Together the two reductions remove most of the excess of the zone mean over the ridge value, so the constant fixed by the local moment lands near the static peak vertex.

The TPSC ansatz itself overestimates the double occupancy relative to the Galitskii--Migdal value, by $9$--$15\%$ at $n \leq 0.6$ and $3$--$7\%$ at $n = 0.7$--$0.8$, i.e., by $2.0$--$6.9$ standard deviations of that value at $T/t \geq 0.10$.
A larger $D$ means a smaller local moment, which the sum rule meets with a smaller vertex; this places $U^\mathrm{TPSC}_\mathrm{sp}$ below $U_\mathrm{eff}$, most at low filling (resolved point by point with $D_{\mathrm{Tr}\Sigma G}$, Appendix~\ref{app:docc}), and accounts for the drift of $U^\mathrm{TPSC}_\mathrm{sp}/U_\mathrm{sp}(\mathbf q^*)$ in Sec.~\ref{sec:peak}.

\section{The self-energy}
\label{sec:sigma}

Section~\ref{sec:vertex} found spin fluctuations that become strongly enhanced toward half filling but remain short ranged.
Strong fluctuations should raise the scattering beyond its second-order value, whereas short-ranged ones should leave its momentum dependence tied to the band energy rather than to particular regions of the Fermi surface.
Both expectations hold at $U/t=4$ for $n = 0.7$--$0.9$: on the $T/t=0.14$ isotherm the nonlocal scattering rate reaches four to five times its second-order value (Sec.~\ref{sec:mechanism}), yet for $T/t \geq 0.14$ it depends almost only on the band energy (Sec.~\ref{sec:structure}).
Section~\ref{sec:channels} identifies the scattering channels, and Sec.~\ref{sec:local} turns to the local self-energy.

\subsection{Growth of the nonlocal scattering rate}
\label{sec:mechanism}

\begin{figure*}[t]
    \centering
    \includegraphics[width=\textwidth]{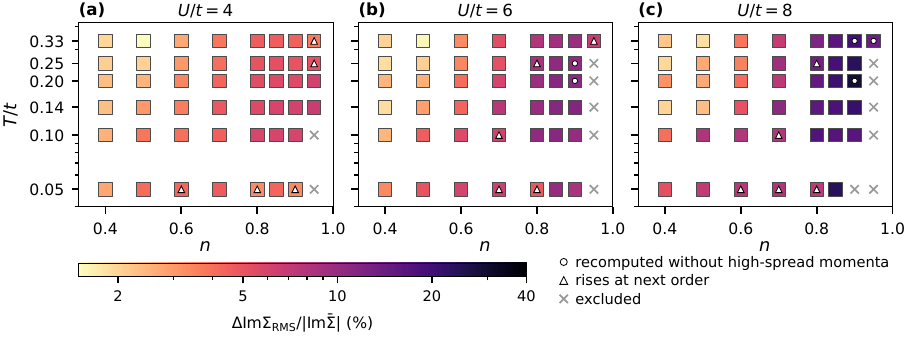}
    \caption{The nonlocal scattering rate stays below about $8\%$ of the local one for $n \leq 0.5$ and grows toward half filling and with $U$.
        Relative momentum dependence of the scattering rate, $\Delta\mathrm{Im}\Sigma_\mathrm{RMS}/|\mathrm{Im}\bar\Sigma|$, over the filling--temperature plane (logarithmic in $T$) at $U/t=4$ (a), $6$ (b) and $8$ (c) for the $131$ retained combinations, on one logarithmic color scale; the truncation term is an estimate rather than a bound (flag $^\mathrm{c}$, Appendix~\ref{app:sys}) at $29$, $35$ and $39$ of the $46$, $43$ and $42$ entries.
        White dots: the five values recomputed without the high-spread momenta, whose $\mathrm{Re}\,\Sigma$ resummation spread exceeds five times the zone median, quoted where this changes $\Delta\mathrm{Re}\Sigma_\mathrm{RMS}/|\mathrm{Re}\bar\Sigma|$ or $\Delta\Sigma_\mathrm{RMS}/|\bar\Sigma|$ by more than two percentage points (tested at $U/t \geq 6$, $n \geq 0.7$); white upward triangles: entries whose value still rises at the next order by more than its truncation term (forward check, Appendix~\ref{app:sys}); gray crosses: combinations excluded as lying at or beyond the magnetic boundary (Sec.~\ref{sec:model}).}
    \label{fig:nonlocality}
\end{figure*}

\begin{figure*}[t]
    \centering
    \includegraphics[width=\textwidth]{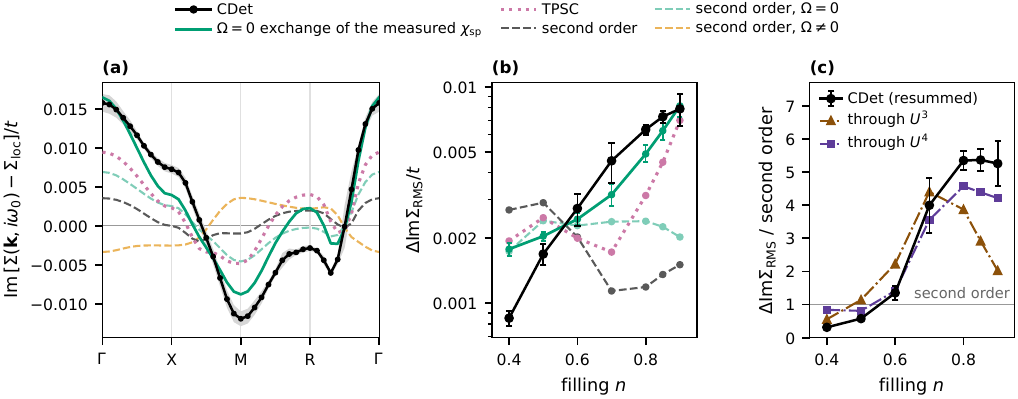}
    \caption{Beyond second order the nonlocal scattering rate grows with filling, and the static exchange of the measured spin fluctuations, $\Sigma_\mathrm{cl}$, follows this growth ($U/t=4$).
        (a) $\mathrm{Im}[\Sigma(\mathbf k,i\omega_0) - \Sigma_\mathrm{loc}]$ along $\Gamma$--$X$--$M$--$R$--$\Gamma$ at $(n=0.8, T/t=0.14)$, each model relative to its own zone mean: CDet (black, band $\pm 1\sigma$); second order (dark-gray dashed) with its $\Omega=0$ term (teal dashed) and $\Omega \neq 0$ remainder (orange dashed); $\Sigma_\mathrm{cl}$, Eq.~\eqref{eq:sigma_cl} (green); TPSC (pink dotted).
        (b) $\Delta\mathrm{Im}\Sigma_\mathrm{RMS}/t$ on the $T/t=0.14$ isotherm for the models of (a) except the $\Omega \neq 0$ remainder; error bars: CDet total uncertainty and the replica $16$--$84\%$ interval of $\Sigma_\mathrm{cl}$.
        (c) $\Delta\mathrm{Im}\Sigma_\mathrm{RMS}$ divided by its second-order value on the isotherm: CDet (resummed, with its systematic uncertainty) and the partial sums of the expansion through $U^3$ (brown) and through $U^4$ (purple); gray line: second order.}
    \label{fig:mechanism}
\end{figure*}

Figure~\ref{fig:nonlocality} shows the nonlocal scattering rate relative to the local one over the whole parameter set: it is at most about $8\%$ for $n \leq 0.5$ at all couplings and grows toward half filling and with $U$.
On the $T/t=0.14$ isotherm at $U/t=4$ the nonlocal scattering rate itself grows by a factor of $9.3(17)$ from $n=0.4$ to $0.9$, whereas its second-order value falls to about half, with a minimum near $n=0.7$ [Fig.~\ref{fig:mechanism}(b)], so that for $n = 0.7$--$0.9$ it is $4.0$--$5.4$ times the second-order value [Fig.~\ref{fig:mechanism}(c)].

Second order fails because its static term, although of the right shape, lacks the spin-fluctuation enhancement.
Split by the bosonic frequency $\Omega$ carried by its bubble, the static ($\Omega=0$) term correlates with the CDet nonlocal part at $0.96$--$0.99$ at the eight susceptibility points with $n \geq 0.7$.
The $\Omega \neq 0$ terms have the opposite momentum structure and cancel $15$--$53\%$ of the static term [Fig.~\ref{fig:mechanism}(a), Appendix~\ref{app:models}].

TPSC writes the self-energy as the exchange of spin and charge fluctuations~\cite{Vilk1997,Moukouri2000},
\begin{align}
    \Sigma(\mathbf k, i\omega_n) - \frac{Un}{2} ={} & \frac{U}{8}\frac{T}{N}\sum_{\mathbf q, m}\bigl[3U_\mathrm{sp}\chi_\mathrm{sp} + U_\mathrm{ch}\chi_\mathrm{ch}\bigr](\mathbf q, i\Omega_m) \nonumber \\
                                                    & \times G_0(\mathbf k+\mathbf q, i\omega_n + i\Omega_m),
    \label{eq:sigma_tpsc}
\end{align}
with $G_0$ at $\mu_0$; with $U_\mathrm{sp} = U_\mathrm{ch} = U$ and $\chi = \chi_0$ it reduces to the second-order term.
As a diagnostic we keep its static spin term with the measured $U_\mathrm{sp}\chi_\mathrm{sp} = 2[R(\mathbf q) - 1]$, the classical ($\Omega=0$) exchange of the measured spin fluctuations,
\begin{equation}
    \Sigma_\mathrm{cl}(\mathbf k, i\omega_n) = \frac{3U}{8}\,\frac{T}{N}\sum_\mathbf{q} U_\mathrm{sp}(\mathbf q)\,\chi_\mathrm{sp}(\mathbf q)\, G_0(\mathbf k+\mathbf q, i\omega_n).
    \label{eq:sigma_cl}
\end{equation}
The data fix only the product $U_\mathrm{sp}\chi_\mathrm{sp}$; the factor $U$ in front of it follows TPSC, and the bare prefactor $U^2\chi_\mathrm{sp}$ in place of $U U_\mathrm{sp}\chi_\mathrm{sp}$ would enlarge $\Sigma_\mathrm{cl}$ by $U/U_\mathrm{sp} \approx 1.6$ without changing its shape or trends.

Spin-fluctuation exchange supplies the missing enhancement.
Relative to its bare-bubble analogue, the static spin term of second order (three quarters of the $\Omega=0$ term in Fig.~\ref{fig:mechanism}(a); Appendix~\ref{app:models}), $\Sigma_\mathrm{cl}$ is enhanced by a factor of $1.8$--$5.4$ for $n \geq 0.7$, which grows with the peak enhancement $R(\mathbf q^*)$.
Its momentum structure correlates with the CDet nonlocal scattering rate at $0.93$--$0.97$ at the eight susceptibility points with $n \geq 0.7$ [Fig.~\ref{fig:mechanism}(a)], and at the seven of them with $T/t \geq 0.10$ it carries $60\%$ to about $100\%$ of the size.
The shape alone is not diagnostic, because the static second-order term has a similar one; the growth is.
On the isotherm $\Sigma_\mathrm{cl}$ grows by a factor of $3.3$ from $n=0.6$ to $0.9$ and the CDet value by $2.9(7)$, while the static second-order term stays flat and full second order is lower at $n=0.9$ than at $n=0.6$ [Fig.~\ref{fig:mechanism}(b)].
The agreement arises at the resummed level, whereas order by order $\Sigma_\mathrm{cl}$ and the exact series differ (Appendix~\ref{app:models}).
The static exchange therefore diagnoses the growth but does not identify the diagrams that produce it, which Sec.~\ref{sec:channels} does.

\subsection{Band-energy dependence and nearly isotropic quasiparticles}
\label{sec:structure}

\begin{figure}[!t]
    \centering
    \includegraphics[width=\columnwidth]{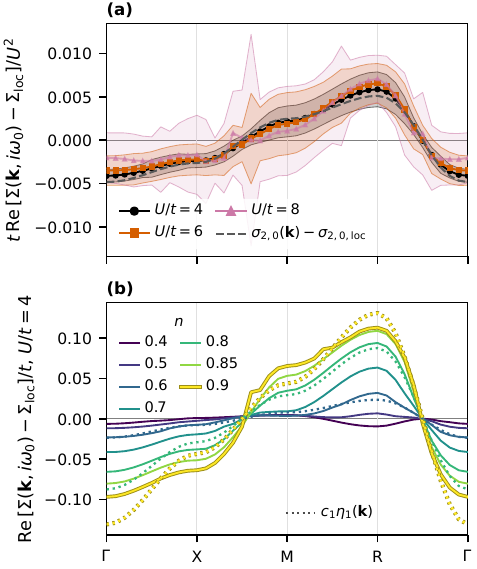}
    \caption{The real part of the nonlocal self-energy keeps the shape of the second-order term at all three couplings at $n=0.8$ (a) and is mainly a renormalization of the nearest-neighbor hopping at $U/t=4$ (b).
        $\mathrm{Re}[\Sigma(\mathbf k,i\omega_0) - \Sigma_\mathrm{loc}]$ along $\Gamma$--$X$--$M$--$R$--$\Gamma$ [Fig.~\ref{fig:map}(b)].
        (a) Divided by $U^2$ at $(n=0.8, T/t=0.14)$ for $U/t=4$ (black circles), $6$ (vermilion squares) and $8$ (reddish-purple triangles) (shaded bands: the systematic uncertainty of the difference); dashed gray: the second-order term $\sigma_{2,0}(\mathbf k) - \sigma_{2,0,\mathrm{loc}}$.
        (b) At $U/t=4$, in units of $t$, on the $T/t=0.14$ isotherm for $n=0.4$--$0.9$ (solid, colored by filling), with its nearest-neighbor harmonic $c_1\eta_1(\mathbf k)$ (dotted, for $n=0.6$, $0.8$, $0.9$).}
    \label{fig:sigma_k}
\end{figure}

At $U/t=4$ the momentum dependence of the self-energy on the Matsubara axis is mainly a dependence on the band energy.
At $(n=0.8, T/t=0.14)$ the nonlocal real part divided by $U^2$ has the shape of the second-order term at all three couplings [Fig.~\ref{fig:sigma_k}(a); correlation over the zone $0.99$, $0.98$ and $0.89$ at $U/t=4$, $6$ and $8$].
It is chiefly a hopping renormalization: at $U/t=4$ the single harmonic $c_1\eta_1(\mathbf k)$ carries $96$--$98\%$ of its variance for $n = 0.6$--$0.9$ on the $T/t=0.14$ isotherm [Fig.~\ref{fig:sigma_k}(b)], and its magnitude relative to second order rises along the isotherm from about $0.3$ at $n=0.5$ through unity at $n \approx 0.8$ to $1.3$ at $n=0.9$.
The renormalized hopping, $t_\mathrm{eff}/t - 1 = -c_1/2t$, reaches $+2.18(8)\%$ at $n=0.9$: the nonlocal self-energy widens the band slightly, against the narrowing by $Z$, as nonlocal exchange does in combined $GW$ and DMFT calculations for a three-dimensional material~\cite{Tomczak2014}.

\begin{figure}[!t]
    \centering
    \includegraphics[width=\columnwidth]{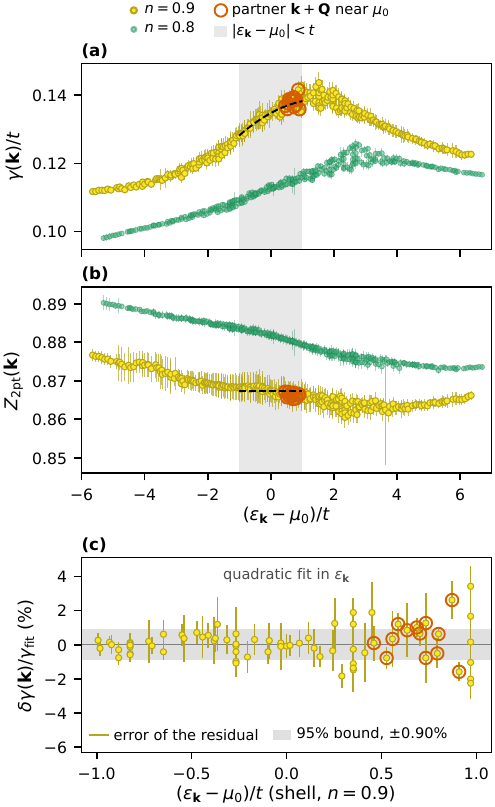}
    \caption{The scattering rate and the quasiparticle weight depend on momentum almost only through the band energy ($U/t=4$, $T/t=0.14$, all momenta at one common order).
        (a) $\gamma(\mathbf k) = -\mathrm{Im}\,\Sigma(\mathbf k, i\omega_0)$ and (b) $Z_\mathrm{2pt}(\mathbf k)$ against $(\varepsilon_\mathbf{k}-\mu_0)/t$ for all $286$ irreducible momenta at $n=0.9$ (yellow) and $0.8$ (green); the per-momentum uncertainties are dominated by a common offset.
        Gray band: the shell $|\varepsilon_\mathbf{k}-\mu_0| < t$; dashed: quadratic fit of $\gamma(\mathbf k)$ and shell mean of $Z_\mathrm{2pt}(\mathbf k)$ at $n=0.9$; vermilion rings: shell momenta at $n=0.9$ whose partner $\mathbf k + \mathbf Q$, $\mathbf Q = (\pi,\pi,\pi)$, lies within $t/4$ of the Fermi level.
        (c) Residuals of $\gamma(\mathbf k)$ from the fit on the $n=0.9$ shell (bars: per-momentum error of the residual); gray band: $\pm 0.90\%$, the $95\%$ upper limit on the root-mean-square variation at fixed band energy; single residuals can lie outside it because of their noise (Appendix~\ref{app:fs}).}
    \label{fig:collapse}
\end{figure}

The imaginary part also depends mainly on the band energy.
On the $T/t=0.14$ isotherm a cubic polynomial in $\varepsilon_\mathbf{k}$ captures $96$--$99\%$ of the zone variance of $\gamma(\mathbf k)$ for $n = 0.7$--$0.85$ and $91\%$ at $n=0.9$, where the rest is mostly non-cubic dependence on $\varepsilon_\mathbf{k}$ (Fig.~\ref{fig:collapse}); at $n \leq 0.6$, where the nonlocal part is small, it captures $59$--$85\%$.

Both parts thus follow the band energy, and the quasiparticles are nearly isotropic (Sec.~\ref{sec:quantities}).
At $(n=0.9, T/t=0.14)$, the most nonlocal point of the isotherm, $\gamma(\mathbf k)$ and $Z_\mathrm{2pt}(\mathbf k)$ of all $286$ momenta lie close to smooth functions of $\varepsilon_\mathbf{k}$ [Fig.~\ref{fig:collapse}(a,b)].
On the shell $|\varepsilon_\mathbf{k} - \mu_0| < t$ at $U/t=4$, $n = 0.7$--$0.9$ and $T/t \geq 0.14$, the scattering rate at fixed band energy varies by at most about $1\%$: the $95\%$ upper limit is $0.9\%$ for the median of the three estimators and up to $1.1\%$ for any single one (Appendix~\ref{app:fs}).
The quasiparticle weight $Z_\mathrm{2pt}(\mathbf k)$ varies by at most $0.14\%$, and at every point by at most $2\%$ of its correlation part $1-Z_\mathrm{2pt}$ (Appendix~\ref{app:fs}).
At lower temperature the common order is too low to bound the variation, although the correlation length stays well below the thermal length there as well.

A small variation remains within these limits, at most a seventh of the nonlocal scattering rate after noise subtraction; its momentum pattern is that of the static exchange $\Sigma_\mathrm{cl}$ of Eq.~\eqref{eq:sigma_cl} (correlation $0.70$--$0.96$), which at $(n=0.9, T/t=0.14)$ predicts $1.2\%$, against the $0.9\%$ upper limit.
For $0.7 \leq n \leq 0.9$ the band energy therefore describes the nonlocal self-energy to good approximation.
At $n \leq 0.6$ the variation at fixed band energy is larger, up to about $2\%$ at $T/t \geq 0.14$; it has the pattern of second order and is a sizable fraction of the small nonlocal part there (Appendix~\ref{app:fs}).

At $T/t \geq 0.14$ hot spots are absent at the percent level, as the short correlation length leads one to expect: $\xi$ stays below half the thermal length in every direction (Sec.~\ref{sec:peak}).
Near half filling, where the spin-fluctuation peak lies at $\mathbf Q = (\pi,\pi,\pi)$, the lattice geometry prevents them.
On the simple-cubic lattice $\varepsilon_{\mathbf k + \mathbf Q} = -\varepsilon_\mathbf{k}$, so exchange of fluctuations at $\mathbf Q$ connects a Fermi momentum to a momentum with band energy $-\mu_0$ rather than to another Fermi momentum, and depends on $\mathbf k$ almost only through $\varepsilon_\mathbf{k}$.
Accordingly, at $(n=0.9, T/t=0.14)$ the momenta whose partner $\mathbf k + \mathbf Q$ lies within $t/4$ of the Fermi level sit $0.46$--$0.91t$ above $\mu_0$, and their scattering rate deviates from the band-energy trend by only $+0.21(17)\%$, within the scatter of the ringed residuals in Fig.~\ref{fig:collapse}(c).
At $n = 0.8$, where the peak lies off $\mathbf Q$, the shell momenta whose partner $\mathbf k + \mathbf q^*$ lies within $t/4$ of the Fermi level include momenta at the Fermi level itself, and they scatter more than the rest of the shell by $0.21(4)\%$, comparable to the $0.13\%$ that static spin-fluctuation exchange predicts (Appendix~\ref{app:fs}).
The spin fluctuations thus leave a resolved imprint at the momenta they connect, but it is a fraction of a percent of the scattering rate.

Nor do the data show the Matsubara-axis signature of a pseudogap: the low-frequency slope, whose reversal would signal one, keeps its sign within its uncertainty at every momentum of the retained points (Appendix~\ref{app:z_freq}).
The pseudogap and marginal-Fermi-liquid behavior that cluster calculations report closer to half filling~\cite{Kellar2023} lie outside what Matsubara data at $T/t \geq 0.05$ resolve~\cite{Song2025PRB}.

For a parametrization of the nonlocal self-energy its frequency dependence matters as well.
At $n=0.7$ and $0.8$ the nonlocal self-energy is close to a product $f(\mathbf k)g(i\omega_n)$ with a single momentum profile $f$ over the first eight Matsubara frequencies (Appendix~\ref{app:fs}), as D$\Gamma$A found at $n = 0.9$--$0.975$~\cite{SchaferTomczak2015}.
For the imaginary part $g$ is not constant: the nonlocal scattering is separable but not frequency independent.

\subsection{Scattering channels and the fermion-boson vertex}
\label{sec:channels}

On the $T/t=0.14$ isotherm with $n = 0.7$--$0.9$, the momentum structure of the nonlocal scattering rate is carried by spin-fluctuation diagrams.
The particle-particle ladder, although it dominates the third-order coefficient, contributes little when summed to all orders and acts mainly through the screening of the spin vertex.
Followed order by order [Fig.~\ref{fig:mechanism}(c)], the growth of the nonlocal scattering rate first appears at third order.
At $n = 0.7$--$0.8$ the partial sum through $U^3$ already reaches about four times the second-order value, with a band-energy dependence nearly orthogonal to the second-order one.
Toward $n=0.9$ the fourth and higher orders take over, whereas at $n \leq 0.5$ successive orders partly cancel and the resummed nonlocal scattering rate stays below its second-order value.

\begin{figure*}[t]
    \centering
    \includegraphics[width=\textwidth]{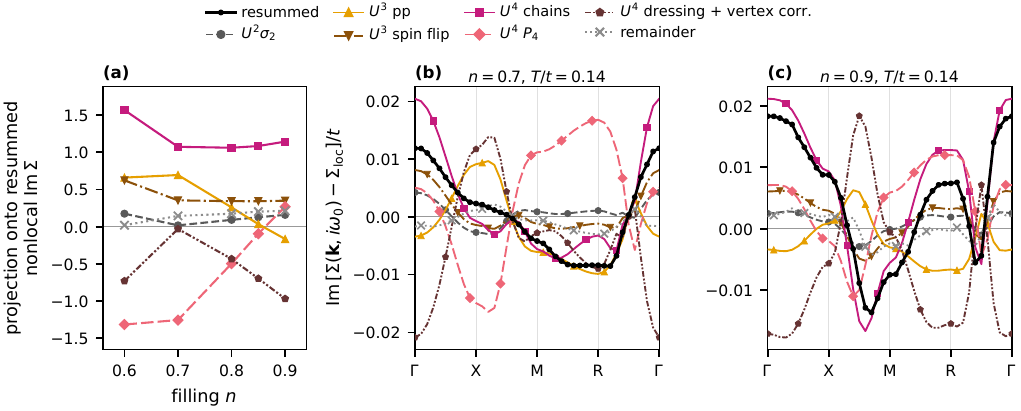}
    \caption{The momentum structure of the nonlocal scattering rate is a sum of large, partly canceling diagram groups ($U/t=4$, $T/t=0.14$).
        (a) Projection of each group onto the resummed nonlocal $\mathrm{Im}\,\Sigma(\mathbf k, i\omega_0)$ on the isotherm (Sec.~\ref{sec:channels}): $U^2\sigma_2$; the third-order particle-particle (pp) and spin-flip ladders; the fourth-order chains, pp ladder $P_4$, and the sum of the dressing and vertex-correction diagrams (the dressing alone projects at most $0.07$); and the remainder beyond fourth order.
        The projections add to $1$ at each filling; $n = 0.4$--$0.5$, where they cancel, are omitted.
        (b), (c) $\mathrm{Im}[\Sigma(\mathbf k, i\omega_0) - \Sigma_\mathrm{loc}]$ along $\Gamma$--$X$--$M$--$R$--$\Gamma$ at $n = 0.7$ and $0.9$: the resummed CDet value (black, all momenta at one common order) and the same pieces, which sum to it; positive values mean less scattering.}
    \label{fig:partial_sums}
\end{figure*}

Exact low-order diagrams on the same lattice identify the channels at third and fourth order (Appendix~\ref{app:h3}).
The third-order coefficient consists of two diagrams built from $G_0$, the particle-particle ladder and the transverse (spin-flip) particle-hole ladder; the latter is the $U^3$ term of spin-fluctuation exchange with the bare interaction, as in the fluctuation-exchange (FLEX) approximation~\cite{Bickers1989,Arita2000}.
At fourth order the measured coefficient is split into three groups: the single-channel ladders (the spin-fluctuation chains and the particle-particle ladder $P_4$) and the self-energy dressing of the sunset, both evaluated exactly, and the vertex-correction diagrams, obtained as the remainder.
The projection of a piece onto the resummed nonlocal scattering rate is the coefficient of its momentum profile along the resummed one.
The projections of all pieces, including the orders beyond the fourth, add to $1$ (Appendix~\ref{app:h3}).

At fourth order the pieces are large and cancel [Fig.~\ref{fig:partial_sums}(a)]: the chains alone project to $1.06$--$1.57$ at $n = 0.6$--$0.9$, more than the whole, and are offset by the particle-particle ladder $P_4$, which dominates at $n \leq 0.7$ and changes sign at $n=0.9$, and by the vertex corrections, which dominate at $n \geq 0.85$.
The self-energy dressing of the sunset, by contrast, projects at most $0.07$, and the orders beyond the fourth only $0.17$--$0.21$ at $n \geq 0.8$ (Appendix~\ref{app:h3}).
The momentum profiles of the pieces show the same cancellation [Fig.~\ref{fig:partial_sums}(b,c)], so a projection gives the sign and weight of a channel along the resummed profile rather than its share of the total.

The groups behave differently from order to order.
The particle-particle ladder makes up most of the third-order coefficient at $n \geq 0.7$ ($68$--$84\%$ of it, Appendix~\ref{app:h3}).
This share says little about the resummed result: the contributions of the ladder alternate in sign, because its series is singular at the pairing (Thouless) pole of the bare ladder, $U \approx -(2.2$--$2.5)t$ on the isotherm, closer to $U=0$ than the physical coupling.
Summed as a $T$-matrix built from $G_0$, it projects at most $0.18$ onto the resummed nonlocal scattering rate at $n = 0.6$--$0.7$ and essentially nothing at $n \geq 0.8$; its main role is the screening of the spin vertex at second order (Sec.~\ref{sec:vertex_sp}).

The spin-fluctuation terms, by contrast, keep their sign, since the singularity of their series, a Stoner pole, lies at positive $U$, and the momentum dependence of the fourth-order chains correlates best with the resummed one ($0.90$--$0.95$).
Built on $G_0$ with the bare interaction, however, that series diverges at $n \geq 0.8$, where the bare Stoner factor exceeds one, so a spin-fluctuation description requires the screened vertex $U_\mathrm{sp} \approx 0.6$--$0.7\,U$ of Sec.~\ref{sec:vertex_sp}.
Dressing the propagators does not remove the requirement: relative to a bubble built from the CDet self-energy, a dressed reference that is not conserving~\cite{Kozik2015}, the zone-mean spin vertex is still $0.70$--$0.84\,U$ at $n \leq 0.8$ (Appendix~\ref{app:kb}).

Static exchange follows the growth of the nonlocal scattering rate, but the full exchange of Eq.~\eqref{eq:sigma_tpsc}, with the measured $\chi_\mathrm{sp,ch}(\mathbf q, i\Omega_m)$ and vertices at $|m| \leq 3$, does not fit the local and the nonlocal scattering together.
It overshoots the local $\gamma$ by $36$--$53\%$ for $n \geq 0.7$ and $T/t \leq 0.14$, whereas its nonlocal scattering rate stays below the CDet value on the isotherm, by $62\%$ at $n = 0.7$ and by $6$--$45\%$ at $n = 0.8$--$0.9$.
A constant fermion-boson vertex $\lambda$ at both ends of the exchanged fluctuation multiplies both parts by $\lambda^2$ and cannot remove this imbalance.
With $\lambda$ fixed by the local $\gamma$, which requires $\lambda^2 = 0.65$--$0.73$, the exchange accounts for between a quarter and two thirds of the CDet nonlocal scattering rate at $n = 0.6$--$0.9$, about half on average at $n = 0.6$--$0.85$.
Other normalizations of the exchange give larger fractions, which reach the CDet value only at $(n=0.9, T/t=0.14)$ (Appendix~\ref{app:models}).
Corrections beyond a constant fermion-boson vertex must therefore suppress the local scattering more than the nonlocal one, and the fourth-order vertex-correction diagrams, which include corrections to that vertex, do so: they remove $66$--$69\%$ of the nonlocal part of the single-channel ladders at $n = 0.6$--$0.9$, but $86$--$90\%$ of their local part.

TPSC limits the overshoot of the local $\gamma$ to $6$--$16\%$ at $n \geq 0.7$ and $T/t \leq 0.14$ because its vertex $U^\mathrm{TPSC}_\mathrm{sp}$ lies about $14\%$ below the zone mean.
It nevertheless shows the same imbalance: its nonlocal scattering rate has a similar shape (correlation $0.68$--$0.89$ for $T/t \geq 0.10$) but $0.38$--$0.88$ of the size, while for $n \geq 0.8$ and $T/t \leq 0.14$ its $c_1$ lies within $12\%$ of the CDet value, inside the CDet uncertainty (Appendix~\ref{app:models}).

\subsection{Local self-energy and quasiparticles}
\label{sec:local}
\label{sec:qp}

A momentum-independent self-energy misses the nonlocal part $\Delta\Sigma_\mathrm{RMS}$ by construction, and its local part can in addition differ from the exact zone average, because the nonlocal correlations it neglects feed back into the local one.
Comparing CDet with paramagnetic single-site DMFT on the same $20^3$ lattice through $Z_\mathrm{2pt}$ and $\gamma$ measures this feedback (Table~\ref{tab:dmft}); at $U/t=4$ it is at most a few percent of $\gamma$, and DMFT captures the low-frequency local $\mathrm{Im}\,\Sigma$ well beyond second order.

At low filling the higher orders suppress the local scattering below its second-order value: at $n=0.4$ the local second-order term overshoots $\gamma$ by $65\%$, and second-order DMFT (IPT-type: the impurity self-energy to second order in the Weiss field, as in iterated perturbation theory) by $30\%$.
DMFT captures the suppression.
It reproduces $Z_\mathrm{2pt}$ within $0.25\%$ at all eight points compared and $\gamma$ within $2.2\%$ for $n \leq 0.7$ (Appendix~\ref{app:dmft}).
The deviation in $\gamma$ is resolved at $n=0.5$; at $n=0.6$ and $0.7$, where it is $1.5\%$ and $0.9\%$, it is smaller than the momentum dependence that DMFT omits, $3.7\%$ and $4.8\%$ of $\mathrm{Im}\,\Sigma(i\omega_0)$.
Where the nonlocal part is this small, single-site DMFT is expected to be accurate, as it is for the energy and density of the strongly doped two-dimensional model~\cite{Kozik2010}.
At $n = 0.8$--$0.9$ and $T/t=0.14$ DMFT lies $1.3$--$3.1\%$ below the CDet $\gamma$, within $1.3$ standard deviations (Appendix~\ref{app:dmft}).
At $U/t=6$ and $8$ no DMFT difference exceeds $1.6$ standard deviations of the larger CDet uncertainties (Table~\ref{tab:dmft}).

\begin{table*}[!tb]
\centering
\footnotesize
\setlength{\tabcolsep}{4pt}
\caption{Single-site DMFT reproduces the CDet local scattering rate and quasiparticle weight at $U/t=4$ within a few percent. Paramagnetic DMFT on the $20^3$ lattice of the CDet runs, with an exact-diagonalization impurity solver and $N_b=6$ bath sites (Appendix~\ref{app:dmft}): values of $Z_\mathrm{2pt}$, Eq.~\eqref{eq:Z}, and $\gamma$ at $U/t=4$, and relative differences $\Delta X = X_\mathrm{DMFT}/X_\mathrm{CDet}-1$ in percent, with the relative CDet uncertainty at $U/t=4$ (systematic and statistical in quadrature) in parentheses in units of the last digit; at $U/t=6$ and $8$ that uncertainty is $0.2$--$8\%$ for $Z$ and $2$--$27\%$ for $\gamma$, and no difference exceeds 1.6 standard deviations. Baselines on the same grid: the local second-order term $U^2\sigma_{2,0}$ and second-order DMFT of iterated-perturbation-theory type (IPT-type), printed to the decimal place of the DMFT entry. $^\sharp$: below the DMFT N\'eel temperature at $U/t=6$ and $U/t=8$.}
\label{tab:dmft}
\begin{tabular}{c cccc cc cc cc cc}
\toprule
 & \multicolumn{8}{c}{$U/t=4$} & \multicolumn{2}{c}{$U/t=6$} & \multicolumn{2}{c}{$U/t=8$} \\
\cmidrule(lr){2-9}\cmidrule(lr){10-11}\cmidrule(lr){12-13}
 & \multicolumn{4}{c}{DMFT} & \multicolumn{2}{c}{second order} & \multicolumn{2}{c}{IPT-type} & \multicolumn{2}{c}{DMFT} & \multicolumn{2}{c}{DMFT} \\
\cmidrule(lr){2-5}\cmidrule(lr){6-7}\cmidrule(lr){8-9}\cmidrule(lr){10-11}\cmidrule(lr){12-13}
$n$ & $Z_\mathrm{2pt}$ & $\Delta Z$ & $\gamma/t$ & $\Delta\gamma$ & $\Delta Z$ & $\Delta\gamma$ & $\Delta Z$ & $\Delta\gamma$ & $\Delta Z$ & $\Delta\gamma$ & $\Delta Z$ & $\Delta\gamma$ \\
\midrule
\multicolumn{13}{l}{$T/t=0.14$} \\
0.40 & 0.9526 & $-0.04(10)$ & 0.0374 & $+2.2(15)$ & $-2.56$ & $+64.9$ & $-1.43$ & $+29.6$ & $0.0$ & $+4$ & $+0.1$ & $+8$ \\
0.50 & 0.9354 & $-0.09(4)$ & 0.0565 & $+2.1(5)$ & $-2.57$ & $+45.5$ & $-1.34$ & $+16.9$ & $-0.3$ & $+3$ & $-0.8$ & $+6$ \\
0.60 & 0.9165 & $-0.1(3)$ & 0.0760 & $+1.5(30)$ & $-2.2$ & $+26.2$ & $-1.1$ & $+7.7$ & $-0.6$ & $+1$ & $-2.0$ & $0$ \\
0.70 & 0.8975 & $0.0(2)$ & 0.0951 & $+0.9(5)$ & $-1.3$ & $+10.4$ & $-0.7$ & $+1.4$ & $+0.1$ & $0$ & $+0.7$ & $-4$ \\
0.80 & 0.8806 & $+0.06(2)$ & 0.1125 & $-1.3(22)$ & $-0.30$ & $-2.6$ & $-0.25$ & $-4.2$ & $+0.3$ & $-4$ & $+0.6$ & $-12$ \\
0.85 & 0.8738 & $+0.1(6)$ & 0.1194 & $-3.1(24)$ & $+0.2$ & $-8.4$ & $-0.1$ & $-7.0$ & $+0.4$ & $-8$ & $+1.4$ & $-17$ \\
0.90$^\sharp$ & 0.8685 & $+0.2(4)$ & 0.1245 & $-2.8(58)$ & $+0.6$ & $-11.3$ & $+0.1$ & $-7.4$ & $+1.2$ & $-5$ & $+5.8$ & $-8$ \\
\midrule
\multicolumn{13}{l}{$T/t=0.05$} \\
0.80 & 0.8303 & $-0.2(6)$ & 0.0408 & $+2.7(46)$ & $-0.3$ & $+2.5$ & $-0.2$ & $-1.1$ & $-0.6$ & $+5$ & $+0.3$ & $+6$ \\
\bottomrule
\end{tabular}
\end{table*}

At $U/t=4$ the quasiparticle weight stays large and is nearly independent of temperature.
At $T/t=0.14$, $Z_1$ falls from $0.925$ at $n=0.4$ to $0.798$ at $n=0.8$ and $0.778$ at $n=0.9$, and for $n \leq 0.8$ it changes by less than $0.02$ between $T/t=0.33$ and $0.05$.
The two-point estimator decreases on cooling by $0.04$--$0.12$ over the same range, but this drift is its curvature bias (Appendix~\ref{app:z_freq}): a linear extrapolation of $Z_\mathrm{2pt}$ to $T=0$ agrees with $Z_1$ at $T/t=0.05$ within $0.008$, and DMFT shows the same difference between the two estimators.
The correlation part $1-Z_1$ at $n=0.8$ is therefore $0.20$, not the $0.12$ that $Z_\mathrm{2pt}$ suggests at $T/t=0.14$.
For $n \leq 0.8$ the lowest $Z_1$ is about $0.80$, $0.62$ and $0.46$ at $U/t = 4$, $6$ and $8$, and $Z_1$ decreases monotonically with filling and coupling; at $U/t = 6$ and $8$ it still decreases toward low temperature at $n \leq 0.7$.

\section{Discussion}
\label{sec:discussion}

At $U/t=4$ and toward half filling, the doped three-dimensional Hubbard model combines strong spin fluctuations with nearly isotropic quasiparticles.
A screened and nearly uniform spin vertex produces a susceptibility peak whose height is set by the slightly smaller vertex on the nesting ridge.
For $n = 0.7$--$0.9$ at $T/t = 0.14$ the fluctuations raise the nonlocal scattering rate to four to five times its second-order value, through spin-fluctuation diagrams rather than the particle-particle ladder, which acts mainly by screening the vertex.
The fluctuations nevertheless remain short ranged, and for $T/t \geq 0.14$ the self-energy depends almost only on the band energy, without hot spots.
Single-site DMFT reproduces the local scattering rate and quasiparticle weight within a few percent.
These results constrain approximate theories at three levels: the spin response of theories with a single vertex, the nonlocal self-energy that extensions of DMFT must add, and the vertices of spin-fluctuation theories.

For theories with a single spin vertex the peak is the most sensitive quantity: that vertex must match the ridge value rather than the zone mean, since the Stoner denominator amplifies any mismatch.
TPSC reproduces the peak because the frequency dependence of the vertex and the reduced high-frequency weight bring its sum-rule constant down close to the ridge value (Sec.~\ref{sec:localmoment}).
Its ansatz for the double occupancy lies $3$--$7\%$ above the Galitskii--Migdal value for $n = 0.7$--$0.8$ and up to $15\%$ above it at low filling, and the local moments of Table~\ref{tab:docc} are a reference for improved schemes~\cite{GauvinNdiaye2023}.

Extensions of DMFT such as D$\Gamma$A or TPSC+DMFT, the latter developed in two dimensions~\cite{Martin2023}, must supply a hopping renormalization of second-order shape and a nonlocal scattering rate that grows with the spin fluctuations.
For $n = 0.7$--$0.9$ and $T/t \geq 0.14$ a parametrization $\Sigma(\varepsilon_\mathbf{k}, i\omega_n)$ captures this nonlocal part to good approximation within $t$ of the Fermi level (Sec.~\ref{sec:structure}), whereas such a parametrization fails in the half-filled two-dimensional model at low temperature~\cite{Pudleiner2016}.
At lower filling the nonlocal part is small at all couplings.

The data constrain spin-fluctuation theories of the self-energy~\cite{Moriya1985} in two ways.
First, exchange of the measured fluctuations through a bare fermion-boson vertex on $G_0$, the form used by TPSC and, on dressed propagators, by FLEX, reproduces the momentum structure of the nonlocal scattering rate (Sec.~\ref{sec:mechanism}) but not its balance with the local part, and no constant rescaling of the vertex restores it (Sec.~\ref{sec:channels}).
In the single-boson-exchange decomposition of the vertex~\cite{Krien2020,Krien2021,Bonetti2022} this balance is set by the fermion-boson (Hedin) vertex and by the multiboson remainder, so either that vertex depends on momentum or frequency, or the multiboson remainder matters.
The fourth-order vertex-correction diagrams, which include corrections to the Hedin vertex, act in the required direction.
A local, frequency-dependent vertex, as in the triply irreducible local expansion (TRILEX)~\cite{Ayral2015}, is one candidate compatible with the local accuracy of DMFT.

Second, on noninteracting propagators the fluctuations must be built on the screened spin vertex (Sec.~\ref{sec:channels}).
Ladder D$\Gamma$A builds its fluctuations on the local vertex of DMFT that is irreducible in the particle-hole channel, a vertex that contains local particle-particle screening~\cite{Toschi2007,Katanin2009}.
Self-consistent dressing, as in FLEX, avoids the divergence of the bare-vertex series without replacing the screening, and its lowest-order term, the dressing of the sunset, projects at most $0.07$ onto the nonlocal scattering rate at $n \geq 0.6$ (Sec.~\ref{sec:channels}).
At half filling in three dimensions such partial summations do not systematically improve the description of the N\'eel transition~\cite{Iskakov2022}.
The vertices of Table~\ref{tab:vertex} allow ladder and functional-renormalization-group schemes~\cite{Toschi2007,Rohringer2016,Katanin2009fRG,Hille2020,Ehrlich2020} to be checked at the level of the vertex.

Two questions remain open.
The first is which part of the fourth-order vertex corrections shifts the balance between local and nonlocal scattering in the required direction.
Separating the corrections to the Hedin vertex from the particle-particle-screened spin-fluctuation exchange and the remaining diagrams would show whether a corrected fermion-boson vertex suffices and whether particle-particle scattering acts on the nonlocal self-energy only through the screening of the spin vertex.
The second concerns stronger coupling: at $U/t \geq 6$ the nonlocal self-energy is known only in its trends (Sec.~\ref{sec:precision}), and near half filling the series have singularities close to the physical coupling (Appendix~\ref{app:singular}), so a stronger-coupling reference at lower temperature is the next step.

\begin{acknowledgments}
    We thank Tu Hong and Yufeng Song for useful discussions.
    P.H. and Y.D. were supported by the National Natural Science Foundation of China (under Grant No.~12275263) and the Quantum Science and Technology--National Science and Technology Major Project (under
    Grant No.~2021ZD0301900). K.C. was supported by the National Key Research and Development Program of China, Grant No.~2024YFA1408604, the National Natural Science Foundation of China under Grants No.~12474245 and No.~12447103, and the GHfund A (202407010637).
\end{acknowledgments}

\appendix

\section{Resummation protocol and error budget}
\label{app:resum}

The resummed values and their uncertainties throughout the paper follow the protocol of this appendix (Secs.~\ref{sec:CDet} and~\ref{sec:precision}).

\subsection{Normalization of the self-energy series}
\label{app:norm}

At $(n=0.6,\beta t=7)$ and $(0.8,7)$ every self-energy series begins at $(j,l) = (2,0)$, and the density-fixed first-order coefficient vanishes identically at all $286$ momenta.
The $(2,0)$ coefficient agrees at every momentum, within the Monte Carlo error, with the second-order sunset diagram evaluated with $G_0$ at $\mu_0$ on the same lattice (root-mean-square pulls $0.24$--$0.61$, median ratios $0.9994$--$1.0001$), whereas a Hartree-insertion tadpole would contribute at the level of $10^{-2}$, four orders of magnitude above the residual difference.
The physical self-energy is therefore $\Sigma = Un/2 + \Sigma_\mathrm{CDet}$, the counterterm being part of the chemical potential, $\mu = \mu_0 + Un/2 - U\alpha(U)$.

\subsection{Estimators, truncation order and the common order}
\label{app:estimators}

\begin{figure*}[t]
    \centering
    \includegraphics[width=\textwidth]{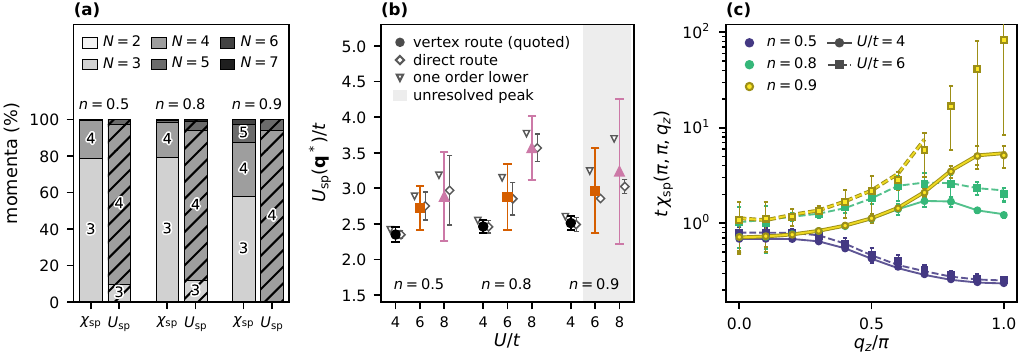}
    \caption{The vertex route stays smooth near the instability, where resumming $\chi_\mathrm{sp}$ first does not; $(n,\beta t) = (0.5,7)$, $(0.8,7)$ and $(0.9,7)$.
        (a) Distribution over the $286$ momenta of the order $N$ accepted by the floor rule for the $\chi_\mathrm{sp}$ series (plain) and the vertex series (hatched), with $N$ written on every segment of at least $9\%$.
        (b) Spin vertex at the peak for $U/t=4$, $6$, $8$: vertex route with its total uncertainty (filled, colored as in Fig.~\ref{fig:sigma_k}), direct route, in which $\chi_\mathrm{sp}$ is resummed first and then inverted (gray open diamonds), and vertex route one order lower (gray down-triangles); gray band: unresolved peak, $U_\mathrm{sp}(\mathbf q^*)\chi_0(\mathbf q^*)/2$ within one standard deviation of unity.
        (c) $\chi_\mathrm{sp}(\pi,\pi,q_z)$ from the direct route (points) and reconstructed from the vertex route (lines) at $U/t=4$ (solid, circles) and $6$ (dashed, squares); at $(n=0.9,\beta t=7,U/t=6)$ the direct values reach $83 \pm 76$ and the reconstruction is blanked within $0.05$ of its pole.}
    \label{fig:control}
\end{figure*}

The partial sums of the truncated series $\sum_{j=2}^{N} a_j U^j$ typically converge by order $5$--$6$ at $U/t=4$, whereas at $U/t=8$ they oscillate and resummation is essential.
Three estimators are applied to the truncated series.
(i) \emph{Pad\'e approximants}: the family of $[L/M]$ approximants with $L+M = N$ or $N-1$, $N$ the applied truncation order, and $|L-M| \leq 2$, discarding members with real poles in $U \in [0,10]$ (poles at $U < 0$, the nearest singularities of the local series, are kept; Appendix~\ref{app:singular}) and outliers farther than five median absolute deviations from the family median; the median of the surviving family is the estimate and half its full range the family spread.
(ii) \emph{DLog-Pad\'e}: the same construction applied to $g(U) = \tfrac{d}{dU}\ln[\Sigma(U)/U^2]$, so that algebraic singularities appear as simple poles of $g$.
(iii) \emph{Borel-Pad\'e}: Pad\'e approximants of the Borel-transformed series $b_j = a_j/j!$, with $\Sigma(U)$ recovered through the Laplace integral, a third continuation with a different analytic structure.
The central value is the median of the converged estimators.

The truncation order is set by the statistical quality of the coefficients.
For each momentum and component the signal-to-noise order is the highest $N$ for which two consecutive orders satisfy $|a_j|/\sigma_{a_j} > 2$; because the coefficients do not depend on the coupling, it is common to all $U$.
Because a downward fluctuation of one coefficient can truncate a momentum far below its neighbors, the order is floored at one less than the zone median of the signal-to-noise orders (the \emph{floor rule}) [Fig.~\ref{fig:control}(a)], per channel, component and bosonic frequency; for $\mathrm{Im}\,\Sigma(\mathbf k,i\omega_0)$ the applied order is typically $N_\mathrm{max}-2$.

Differences between momenta need one more step.
At $U/t=4$ the truncation error of $\Sigma(\mathbf k, i\omega_0)$ is nearly the same at every momentum: at $(n=0.9, T/t=0.14)$ the local scattering rate changes by $13\%$ between orders $5$ and $7$, about twice the whole nonlocal part.
Momenta truncated at different orders therefore acquire spurious relative shifts, and the nonlocal quantities, built from $\Sigma(\mathbf k, i\omega_n) - \Sigma_\mathrm{loc}(i\omega_n)$, are evaluated with all momenta at one common order per $(n,\beta)$ and component, the $O_h$-weighted median of the floor-rule orders; local quantities keep the floor-rule orders.
Where the common order lies on the plateau of the series, the error is common to the zone and cancels in the difference; a forward check identifies the points where it does not (Appendix~\ref{app:sys}).
At $(n=0.9, T/t=0.14)$ the common order is $7$, and $\Delta\mathrm{Im}\Sigma_\mathrm{RMS} = 0.00790t$ at orders $7$ and $8$ alike, against $0.0094t$ with the floor-rule orders.

The sensitivity to the truncation order is quantified by repeating the entire analysis with the order of every series lowered by one (not below order $3$ for the self-energy and $2$ for the susceptibility and vertex series); the change $\delta_1 = |x(N) - x(N-1)|$ of a quantity $x$ is its one-step truncation term.

\subsection{Systematic uncertainty}
\label{app:sys}

Each reported quantity carries the resummation budget
\begin{align}
    \Delta_\mathrm{sys} = \max\Bigl( & \max_e s_e,\; \max_e |x_e - \tilde{x}|, \nonumber      \\
                                     & \tfrac{1}{2}\bigl[\max_e x_e - \min_e x_e\bigr]\Bigr),
    \label{eq:error_budget}
\end{align}
where $e$ runs over the converged estimators, $x_e$ is the estimate of method $e$, $s_e$ its within-family spread, and $\tilde x$ the multi-method median.
For a zone-derived quantity such as $Z_\mathrm{2pt}$, $\gamma$ or $\Delta\Sigma_\mathrm{RMS}$ the per-momentum budgets are propagated to the zone average as independent errors, which suppresses the part of the estimator disagreement that is coherent over momenta; such a quantity therefore also carries the between-estimator spread of the quantity itself, $\Delta_\mathrm{method} = \max_m |x_m - x|$, with $x_m$ obtained when estimator $m$ is used at every momentum (on the momenta common to the estimators where one of them fails).
For momentum differences such as $\Sigma(\mathbf k,i\omega_0) - \Sigma_\mathrm{loc}(i\omega_0)$ (Fig.~\ref{fig:sigma_k}) the budget is evaluated on the difference itself, each estimator being applied at every momentum before the local projection is subtracted.

\emph{Two-step check and enlarged truncation term.}
The truncation term is the largest part of the systematic uncertainty of the self-energy, and as a one-step shift it bounds the truncation error only if successive shifts shrink fast enough; the analysis was therefore repeated with every series lowered by two orders and the step ratio $r = [x(N) - x(N-1)]/[x(N-1) - x(N-2)]$ formed at the retained points.
For $\gamma$ at $U/t=4$, $r \geq 0.5$, so that a geometric remainder exceeds $\delta_1$, at $24$ of the $46$ points, mainly at $n=0.6$ and at $\beta t \geq 10$, where the series are shortest; for $\Delta\mathrm{Im}\Sigma_\mathrm{RMS}$ at the common order $|r| \geq 1$ at $18$ points, mostly where the nonlocal part is small.
For $\gamma$, $\Delta\mathrm{Im}\Sigma_\mathrm{RMS}$ and $U_\mathrm{sp}(\mathbf q^*)$ the truncation term is enlarged to
\begin{equation}
    \Delta_\mathrm{trunc} =
    \begin{cases}
        \delta_1,                             & -1 < r < 0.5,                     \\
        \max(\delta_1, \delta_\mathrm{tail}), & 0.5 \leq r < 0.8,                 \\
        \max(\delta_1, \delta_2),             & r \geq 0.8 \text{ or } r \leq -1,
    \end{cases}
    \label{eq:trunc_rule}
\end{equation}
with the geometric tail $\delta_\mathrm{tail} = |r/(1-r)|\,\delta_1$, which diverges as $r \to 1$ and is therefore replaced by the two-step shift $\delta_2 = |x(N) - x(N-2)|$ for $r \geq 0.8$; where the $N-2$ level carries no information the term is $\delta_1$.

A backward check cannot detect a common order that lies below the plateau of the series, as at $\beta t = 20$ near half filling, where the odd coefficients are small: at $(n=0.8, \beta t=20)$, $\Delta\mathrm{Im}\Sigma_\mathrm{RMS}$ is $0.00157t$, $0.00131t$ and $0.00133t$ at orders $3$, $4$ and $5$, the common order, but $0.00230t$ and $0.00243t$ at orders $6$ and $7$.
A forward check therefore compares each value with the one at order $N+1$, all momenta at that order; for $\gamma$ the floor-rule value is compared with the zone average at the common order $N+1$, and from a root-mean-square indicator the coefficient noise of its order-$(N+1)$ value, estimated by Gaussian resampling, is subtracted first.
Where the step exceeds both the truncation term and twice its noise, the term is widened to cover it, and for zone averages also the step to order $N+2$.

An entry is flagged (marked $^\mathrm{c}$) where $r \geq 0.8$ or $r \leq -1$, where the forward check widens the term, or where the truncation term exceeds the value: there the truncation term is an estimate rather than a bound.
The flag marks $\gamma$ at $20$, $20$ and $23$ and $\Delta\mathrm{Im}\Sigma_\mathrm{RMS}$ at $29$, $35$ and $39$ of the retained points at $U/t = 4$, $6$ and $8$; at $U/t=4$, $\Delta\mathrm{Im}\Sigma_\mathrm{RMS}$ is unflagged at $14$ of the $16$ points with $n = 0.7$--$0.9$ and $T/t \geq 0.14$, among them the four on the $T/t=0.14$ isotherm.

Every systematic uncertainty is $\sqrt{\Delta_\mathrm{sys}^2 + \Delta_\mathrm{method}^2 + \Delta_\mathrm{trunc}^2}$, where the middle term applies to zone-derived quantities only and $\Delta_\mathrm{trunc}$ is $\delta_1$ or, for $\gamma$, $\Delta\mathrm{Im}\Sigma_\mathrm{RMS}$ and $U_\mathrm{sp}(\mathbf q^*)$, the term of Eq.~\eqref{eq:trunc_rule}, forward-checked for the first two; the statistical error is quoted separately.

\subsection{Statistical uncertainty and replication}
\label{app:stat}

The statistical error comes from a thread-level bootstrap carried through the complete analysis chain: the per-thread measurement files of a run are resampled with replacement, and each of the $100$ replicas passes through the density fixing, the order truncation and the multi-method resummation with the same rules as the central analysis, so that fluctuations correlated across orders and momenta are preserved.
The quoted $\sigma_\mathrm{stat}$ is the half-width of the central $68\%$ replica interval (for the finite-frequency vertices and the sum-rule $D$, the standard deviation of $30$ Gaussian coefficient replicas); for the vertices, whose error bars combine both, it is added to the systematic in quadrature.

The point $(n=0.7,\beta t=7)$ was sampled in two independent runs with $N_\mathrm{max} = 7$ and $8$; the two determinations of $Z_\mathrm{2pt}$ agree to $0.04\%$, $0.17\%$ and $0.35\%$ at $U/t = 4$, $6$ and $8$, and $\gamma$ to $0.3\%$, $0.8\%$ and $1.8\%$; with both runs at the same common order, $\Delta\mathrm{Im}\Sigma_\mathrm{RMS}/|\mathrm{Im}\bar\Sigma|$ reproduces to within $0.4\%$ of its value at all three couplings.

\subsection{Series singularities}
\label{app:singular}

The coefficients of $\gamma$ alternate in sign at $47$ of the $48$ points, and the nearest stable DLog-Pad\'e singularity of the local series lies on the negative real axis at $45$ of them, at a median $7t$ ($4$--$11t$) for $n = 0.5$--$0.9$, and at $n=0.6$ it stays at $|U_s| \approx 4.7$--$5.1t$ for $\beta t = 3$--$10$.
Positive-axis singularities lie at $5.5$--$7.1t$ for $n = 0.6$--$0.7$ and at $8$--$13t$ for $n \geq 0.85$, in fewer approximants; near half filling they make the $U/t=8$ results qualitative.
The negative-axis singularity belongs to the particle-particle ladder, whose bare pairing pole lies at $-1/\Phi(\mathbf 0, 0) = -(1.6$--$3.4)t$ over the parameter set, so that for $T/t \lesssim 0.2$ the formal radius of convergence lies below $U = 4t$.
On the $20^3$ lattice, however, the zero-momentum pair mode carries half of the ladder's coefficient only from order $11$ at $\beta t = 20$, beyond the computed orders, and through them the continuum of pair momenta sets an effective radius of about $5t$; the resummed values thus continue the series through a singularity that the computed orders do not resolve.
A conformal map that opens the negative real axis, $w = [\sqrt{1-U/U_s} - 1]/[\sqrt{1-U/U_s} + 1]$ with $U_s = -5t$, restores the one-step bound for $\gamma$ at $10$ of the $12$ points at $U/t=4$ whose shifts do not shrink ($|r| \geq 1$); the mapped values agree with the tabulated ones within half a systematic uncertainty, except at $\beta t = 20$ and $n \geq 0.8$, where they lie up to two systematic uncertainties higher.

\section{Systematic checks}
\label{app:checks}

This appendix documents the exclusion of combinations near magnetic order (Sec.~\ref{sec:model}) and the finite-size effects (Sec.~\ref{sec:precision}).

\subsection{Phase classification}
\label{app:phase}

Because one run serves all three couplings, the paramagnetic condition is checked per $(n,\beta,U)$ against the critical couplings $U_c(n,T)$ of Ref.~\cite{Lenihan2022}, read from its Fig.~3 ($T/t = 0.05$--$0.3$, $n \geq 0.89$) and interpolated linearly in $T$; the thirteen combinations at which $U$ is not below $U_c$ within its uncertainty, all at $n \geq 0.9$, are excluded.
At $T/t = 0.2$ and $0.25$ the $U_c(n)$ curves terminate at $n^* \approx 0.93$ and $0.95$, below which no order occurs at any $U$, and the incommensurate branch at $T/t=0.05$ reaches $n \approx 0.89$ only at $U \approx 8t$, so the $n=0.9$ points at $T/t \geq 0.10$ are expected to lie outside the ordered region at all three couplings.
For $n \leq 0.85$ at $U/t = 6$, $8$ no controlled boundary exists (D$\Gamma$A places the magnetic quantum critical point near $n \approx 0.8$ at $U \approx 9.8t$~\cite{Schafer2017}).
At $(n=0.9, \beta t=7)$ the paramagnetic series of the static spin susceptibility has a singularity on the positive axis at $U \approx 5.3$--$6.8t$, and at $U/t=6$ and $8$ the Stoner factor of the peak vertex, $U_\mathrm{sp}(\mathbf q^*)\chi_0(\mathbf q^*)/2 = 1.03(21)$ and $1.04(33)$, lies within one standard deviation of unity, so the peak there cannot be distinguished from a divergence (Sec.~\ref{sec:precision}). At $U/t=4$ the measured enhancement $R(\mathbf q^*)$ is finite at all twelve susceptibility points (Table~\ref{tab:peak}), as the boundary requires.

\subsection{Finite-size effects}
\label{app:finite_size}

Four parameter points spanning the sampled regime, $(n,\beta t) = (0.4,3)$, $(0.6,5)$, $(0.6,10)$ and $(0.8,7)$, recomputed on an $L=24$ lattice ($455$ inequivalent momenta), reproduce $Z_\mathrm{2pt}$ to within $0.3\%$ at $U/t \leq 6$ and $0.8\%$ at $U/t=8$, and $\gamma$ to within $1\%$ and $2.6\%$.
Both lattices select the same common order of $\mathrm{Im}\,\Sigma(\mathbf k, i\omega_0)$ at all four points, and at that order $\Delta\mathrm{Im}\Sigma_\mathrm{RMS}$ agrees within $0.35$ of the combined uncertainty at every point and coupling, and $\Delta\mathrm{Re}\Sigma_\mathrm{RMS}$ within $0.27$.
At the most nonlocal point, $(n=0.8,\beta t=7)$, the $L=24$ values of $\Delta\mathrm{Im}\Sigma_\mathrm{RMS}$ are $3\%$, $6\%$ and $4\%$ larger at $U/t = 4$, $6$ and $8$; the largest difference in units of the statistical error alone, $2.1$ standard deviations at $U/t=4$, is smaller than the systematic uncertainty of that entry, $5\%$, so the nonlocal budget carries no separate finite-size term.
The exact size dependence of the local second-order term ($L=20$ against $L \leq 200$) is below $1.3\%$ in $\gamma$ for $\beta t \leq 10$, but $-2.7\%$ to $+4.3\%$ at $\beta t=20$: there the values of $\gamma$, whose quoted uncertainty refers to the $L=20$ lattice, can differ from the thermodynamic limit by about that uncertainty, while the comparison with DMFT on the same lattice is unaffected (Appendix~\ref{app:dmft}).

For the susceptibilities the exact size dependence of the bare bubble ($L=20$ against $L \leq 100$), inserted into Eq.~\eqref{eq:vertex_def} with the measured vertex, changes $\chi_\mathrm{sp}(\mathbf q^*)$ at $U/t=4$ by at most $2.3\%$ at $\beta t \leq 10$ and lowers $\chi_\mathrm{sp}(\mathbf q^*)$ and $R(\mathbf q^*)$ by $9.8\%$ and $7.3\%$ at $(n=0.8,\beta t=20)$, against the ${\approx}29\%$ uncertainty of that point.
A second susceptibility run on $L=24$ at $(n=0.8,\beta t=7)$ gives the same static vertex within its uncertainty, $0.03(3)$ apart at the ten momenta common to the two grids, and rules out vertex changes larger than about $6\%$ there.

\section{The spin vertex and the spin-fluctuation peak}
\label{app:vertices}

This appendix supports Sec.~\ref{sec:vertex}: the vertex route and the precision of the vertices, their screening, the peak vertex and peak line, the correlation length and peak position, and the sum rules.

\subsection{The vertex route and the precision of the vertices}
\label{app:vertex_route}

Near the instability $\chi$ is close to a pole, which every Pad\'e-type estimator resolves badly, whereas the vertex is a smooth function of $U$ there [Fig.~\ref{fig:control}(b,c)].
Two internal checks confirm the order-by-order inversion of Eq.~\eqref{eq:vertex_inv}: the zeroth-order coefficient of the measured static $\chi$ series equals the lattice Lindhard value of Eq.~\eqref{eq:lindhard} to four digits at every momentum, and the first-order coefficient of the vertex series, exactly $1$ since $U_\mathrm{sp} \to U$ at weak coupling, lies between $0.993$ and $1.019$ over all twelve points, both channels and all $286$ momenta (median $0.9996$).

At $U/t=4$ the total uncertainty of $U_\mathrm{sp}(\mathbf q)$ has a median of $2.5\%$ over the $3431$ cells (all momentum-point pairs but one without a converged estimator), $4$--$7\%$ at the four points with $T/t \leq 0.10$, and $7\%$ for the charge vertex, against $6\%$ and $12\%$ for the spin vertex at $U/t=6$ and $8$.
The truncation shift has the same sign at $95$--$96\%$ of the momenta, so averaging does not reduce it, and the zone mean $\langle U_\mathrm{sp}\rangle_\mathbf q$ has a median uncertainty of $3.0\%$, $6.7\%$ and $12.6\%$ at $U/t = 4$, $6$ and $8$.
Cells whose resummation and statistical uncertainties exceed $30\%$ of the value are left out of the figures and zone means ($9$ charge cells and no spin cell at $U/t=4$).
The slopes $B$ of Table~\ref{tab:vertex}, from zone-weighted least squares, are not produced by the truncation: two orders lower $B$ is smaller at ten of the twelve points.

\subsection{Screening of the spin vertex}
\label{app:kb}

The combination $1/\langle U_\mathrm{sp}\rangle_\mathbf q - 1/U$, which the Kanamori--Brueckner form predicts to be the constant $1/U_\mathrm{K}$, rises only from $0.116/t$ to $0.127/t$ between $U/t=4$ and $8$, but it rises at every point, so the form with $U_\mathrm{K} = 8.2t$ is a first-order description rather than an exact one.
The independent estimate uses the two-particle $T$-matrix at the band bottom, $U/(1 + U\Lambda)$ with $\Lambda = W_\mathrm{sc}/4t$ and $W_\mathrm{sc} = 0.5055$ the simple-cubic Watson integral, i.e., $U_\mathrm{K} = 1/\Lambda = 7.9t$.
No Kanamori--Brueckner variant, including one with the particle-particle propagator of the $20^3$ lattice in place of $\Lambda$, follows the filling dependence of the measured vertices.

\emph{Second order by channel.}
Writing $U_\mathrm{sp}(\mathbf q) = U + a_2(\mathbf q)U^2 + O(U^3)$, Eq.~\eqref{eq:vertex_inv} gives $a_2 = 2[\chi^{(2)} - \chi_0^3/4]/\chi_0^2$, with $\chi^{(2)}$ the second-order coefficient of the static $\chi_\mathrm{sp}$.
Beyond the random-phase chain, the second-order diagrams of $\chi_\mathrm{sp}$ are the self-energy insertions and the density-fixing counterterm on the bubble lines (together, the dressing), the particle-particle diagram, and two particle-hole vertex corrections, the crossed transverse one in $\chi_{\uparrow\downarrow}$ and the direct longitudinal one in $\chi_{\uparrow\uparrow}$.
The latter two are the same integral and cancel exactly in $\chi_\mathrm{sp} = 2(\chi_{\uparrow\uparrow} - \chi_{\uparrow\downarrow})$, so $a_2$ is the particle-particle term plus the dressing.
Evaluated exactly at the seven $T/t = 0.14$ points, the sum reproduces $a_2$ at all $286$ momenta (pull root mean square $0.39$--$0.60$); the particle-particle term gives $65$--$78\%$ of its zone mean and, at $n \geq 0.7$, the whole slope against $\chi_0$ (correlation $-0.95$ to $-0.99$).

\emph{Dressed reference.}
Equation~\eqref{eq:vertex_inv} measures the vertex relative to the bare bubble, so self-energy corrections to the bubble are part of $U_\mathrm{sp}(\mathbf q)$.
Referenced instead to the bubble built from the resummed CDet self-energy of the same point, which is available for $n \leq 0.8$ at $T/t \geq 0.10$ and is not a conserving reference~\cite{Kozik2015}, the zone-mean spin vertex rises to $2.78$--$3.36t$, i.e., $0.70$--$0.84\,U$, so dressing does not remove the screening.
The slope $B$ keeps its sign and significance at seven of the eight points and is larger than the bare one at the four with $n \geq 0.7$ [at $(n=0.8,\beta t=7)$, $1.75$ against $1.30$, fitted as in Table~\ref{tab:vertex}]; only at $n=0.4$ does it become consistent with zero, $1.42(29) \to 0.05(30)$ (statistical errors).

\subsection{Peak vertex and peak line}
\label{app:zone_err}

At the peak momentum the vertex series retain four orders at nine of the twelve points, where their terms alternate in sign and, from second order on, shrink by factors of $0.23$--$0.45$ per order and the peak vertex converges monotonically from above; its quoted uncertainty is $3.7$--$6.1\%$ there, $15$--$19\%$ at the two three-term points, and $0.25\%$ at $(n=0.9,\beta t=3)$, where six orders are retained.
A geometric (Aitken) extrapolation of these series lies $2.8\%$ below the applied-order value on average; a downward shift of about this size, common to all points and therefore not constrained by their scatter, is included in the uncertainties of $\bar U$ and of $U_\mathrm{eff}/U_\mathrm{sp}(\mathbf q^*)$ (below and Appendix~\ref{app:docc}).
Along the truncation sequence the median of $U_\mathrm{sp}^\mathrm{TPSC}/U_\mathrm{sp}(\mathbf q^*)$ rises from $0.872$ two orders lower, where the vertex series is essentially of second order, to $0.976$ at the applied order: the agreement of Sec.~\ref{sec:peak} grows with order.

The common vertex $\bar U$ of the peak line is the weighted mean of the direct-route peak vertices $2[1 - 1/R(\mathbf q^*)]/\chi_0(\mathbf q^*)$, so that the line and the plotted $1/R$ come from the same resummation.
At $U/t=4$ these peak vertices lie between $2.35t$ and $2.49t$, a root-mean-square spread of $1.8\%$, and agree with the vertex-route values of Table~\ref{tab:vertex} within their uncertainties.
With the per-point uncertainties treated as independent, $\bar U = 2.431(23)t$ with $\chi^2/\mathrm{dof} = 0.18$: these uncertainties are largely common to the points.
The quoted $\bar U = 2.43(7)t$ therefore combines in quadrature the independent part that the scatter requires ($0.010t$, a fraction $0.37$ of the quoted systematic), the coherent difference between the estimators ($0.031t$), the one-order truncation shift ($0.002t$) and a common truncation tail of the size of the mean Aitken shift at the same peaks ($0.06t$, above).

\subsection{Correlation length and peak position}
\label{app:corr}

The Ornstein--Zernike fit of Sec.~\ref{sec:vertex_def} uses the momenta within three grid steps of the peak.
Across the ridge $\xi$ is $0.41$--$1.39$ lattice spacings (median $0.67$), largest at $(n=0.9,T/t=0.14)$ with $1.39(23)$; along the ridge the background-subtracted width of the Lorentzian fit that locates $q_z^\mathrm{fit}$ in Table~\ref{tab:peak} gives $0.8$--$2.1$.
That fit, a Lorentzian plus a constant, uses the eleven ridge values; the uncertainty of $q_z^\mathrm{fit}$ is the largest of the parametric-bootstrap error, the $16$--$84\%$ half-width over the thread-bootstrap replicas of the ridge, and the half-width of the $q_z$ interval over which the fitted $\chi_\mathrm{sp}$ lies within one standard deviation of its maximum.
The thermal length uses the Fermi-surface average of $|\nabla\varepsilon_\mathbf{k}|$ of the continuum band at $\mu_0$, $\langle v_F\rangle = 2.21$--$2.62\,t$ in units of the lattice spacing; $\xi_\mathrm{th}$ is shortest, $2.5$, at $T/t=0.33$, where $\xi/\xi_\mathrm{th}$ is largest, $0.33$ across and $0.47$ along the ridge.

\subsection{Sum rules, double occupancy and the sum-rule vertex}
\label{app:docc}

The TPSC spin sum rule is dominated by momenta and frequencies away from the peak: at $U/t=4$ and $T/t \leq 0.14$ the static term contributes $6$--$16\%$ of its value and the momenta near the peak $1$--$5\%$, and with the measured $\chi_\mathrm{sp}(\mathbf q, i\Omega_m)$ the computed frequencies $|m| \leq 3$ carry $85$--$93\%$ of its sensitivity to the vertex at the $T/t=0.14$ points and at least $65\%$ at all twelve.

\emph{Double occupancy.}
Equation~\eqref{eq:sumrules} gives $D = (S_\mathrm{ch} - S_\mathrm{sp} + n^2)/4$, with $S_\mathrm{sp/ch}$ the two frequency--momentum sums; in this difference the mismatch of the $1/\Omega^2$ tail amplitude cancels.
The sums are evaluated with the measured $\chi_\mathrm{sp/ch}(\mathbf q,i\Omega_m)$ at $|m| \leq 3$ (vertex route) and, for $|m| \geq 4$, with the random-phase form of Eq.~\eqref{eq:vertex_def} on the exact lattice Lindhard function and the per-momentum vertex frozen at its $m=3$ value.
The tail systematic, the spread of five tail models (the frozen vertex, the zone-median $m=3$ vertex, the bare $U$, the TPSC constants, and a bubble rescaled to satisfy the Pauli sum rule $S_\mathrm{sp} + S_\mathrm{ch} = 2n - n^2$), is $0.4$--$9\%$ of $D$, and the total relative uncertainty of $D$ is $3$--$12.5\%$.
The construction reproduces the first-order coefficient of $D$ to $-0.4$ to $+0.8\%$, and the frozen and bare-$U$ tails bracket the share of $|m| \geq 4$ that the exact second-order coefficient fixes.

\emph{Galitskii--Migdal cross-check.}
Independently of the susceptibilities, $D$ follows from the self-energy,
\begin{equation}
    UD = \frac{Un^2}{4} + \frac{T}{N}\sum_{\mathbf k, \omega_n}\Bigl[\Sigma(\mathbf k, i\omega_n) - \frac{Un}{2}\Bigr]\, G(\mathbf k, i\omega_n),
    \label{eq:gm}
\end{equation}
with $G$ built from the resummed $\Sigma$ at the chemical potential $\mu_0 - U\alpha(U)$, since $\Sigma - Un/2 = \Sigma_\mathrm{CDet}$ (Appendix~\ref{app:norm}); beyond the eight exported frequencies the sum uses the exact second-order sunset plus a remainder decaying as $\omega_n^{-2}$.
This construction reproduces the exact first- and second-order coefficients of $D$ at all twelve susceptibility points, whereas assigning $U\alpha$ to $\Sigma$ would make the first-order coefficient too small in magnitude by $10$--$42\%$.
At the nine points with $n \leq 0.8$, where $\Sigma$ was resummed at all eight frequencies, $D_{\mathrm{Tr}\Sigma G}$ (uncertainty $0.0007$--$0.0016$) agrees with the sum-rule $D$ (Sec.~\ref{sec:localmoment}), validating the modeled tail of the sum rule, and at the eight with $T/t \geq 0.10$ it is the reference for the comparisons of the text.
At $(n=0.8, T/t=0.05)$, where $\gamma$ is flagged and rises by about $4\%$ at the next order, the uncertainty of $D_{\mathrm{Tr}\Sigma G}$ contains only the one-step truncation term.

\emph{Sum-rule vertex.}
$U_\mathrm{eff}$ is the constant vertex whose random-phase local sum on the bare bubble equals $n - 2D$ with the sum-rule $D$, i.e., TPSC without the ansatz.
Over the twelve points $U_\mathrm{eff}/U_\mathrm{sp}(\mathbf q^*) = 1.015$--$1.074$, with inverse-variance mean $1.027(18)$ for independent points and $\chi^2/\mathrm{dof} = 0.23$; adding the possible common shift of the peak vertex (Appendix~\ref{app:zone_err}) gives the $1.03(4)$ of Sec.~\ref{sec:localmoment}.
With $D_{\mathrm{Tr}\Sigma G}$ at the eight points with $n \leq 0.8$ and $T/t \geq 0.10$, $U_\mathrm{eff}/U_\mathrm{sp}(\mathbf q^*) = 1.01$--$1.05$ and $U^\mathrm{TPSC}_\mathrm{sp}/U_\mathrm{eff} = 0.89$--$0.97$, $2.1$--$7.6$ standard deviations below unity, so the drift of $U^\mathrm{TPSC}_\mathrm{sp}/U_\mathrm{sp}(\mathbf q^*)$ carried by the ansatz is resolved point by point.
The ratio $U_\mathrm{eff}/\langle U_\mathrm{sp}\rangle_\mathbf q = 0.89$--$0.94$ is the product of the two reductions of Sec.~\ref{sec:localmoment}.
The rescaling of the Pauli-sum-rule tail model gives a kinetic energy of $0.92$--$0.99$ of its noninteracting value at the eight points with DMFT data, within $0.02$ of the DMFT ratio.

\section{Self-energy analyses}
\label{app:sigma}

This appendix supports Sec.~\ref{sec:sigma}: the fits and correlations, the model self-energies, the Fermi-surface bounds and the frequency dependence, the scattering channels, and the comparison with DMFT and the quasiparticle weight.

\subsection{Fits and correlations}
\label{app:pt2}

A ratio to second order (Sec.~\ref{sec:quantities}) uses the same $O_h$ weights in both indicators and carries the systematic of the resummed one alone.
The shape correlation is the $O_h$-weighted Pearson correlation over the zone between two nonlocal parts.
$c_1$ comes from a weighted least-squares fit of $\mathrm{Re}[\Sigma(\mathbf k,i\omega_0) - \bar\Sigma]$ to $\eta_1(\mathbf k)$; its uncertainty combines the propagated per-momentum budgets, the between-estimator spread and the truncation term.
The band-energy fractions of Sec.~\ref{sec:structure} are the shares of the weighted zone variance of $\gamma(\mathbf k)$ captured by a weighted cubic polynomial in $\varepsilon_\mathbf{k} - \mu_0$.

\subsection{Spin-fluctuation self-energies}
\label{app:models}

The TPSC self-energy, Eq.~\eqref{eq:sigma_tpsc}, is the second step of Ref.~\cite{Vilk1997} in the crossing-symmetric form of Ref.~\cite{Moukouri2000}, with $G_0$ at the density-fixed $\mu_0$ of the run; with $U_\mathrm{sp} = U_\mathrm{ch} = U$ and $\chi = \chi_0$ the Matsubara convolution reproduces the second-order term to $5\times 10^{-5}$.
The static spin term of second order is Eq.~\eqref{eq:sigma_cl} with $U\chi_0$ in place of $U_\mathrm{sp}\chi_\mathrm{sp}$, three quarters of the full static ($\Omega=0$) term of second order shown in Fig.~\ref{fig:mechanism}; the $\Omega \neq 0$ terms have the opposite momentum structure in part because the $m=-1$ term evaluates $G_0$ at $-i\pi T$, where $\mathrm{Im}\,G_0$ changes sign.
Every model is compared with the CDet nonlocal $\mathrm{Im}\,\Sigma(\mathbf k, i\omega_0)$ through the shape correlation and the ratio of $\Delta\mathrm{Im}\Sigma_\mathrm{RMS}$, each about its own zone mean; at $(n=0.8, T/t=0.05)$ the common order lies below the plateau (Appendix~\ref{app:sys}) and no size ratio is quoted.
Order by order $\Sigma_\mathrm{cl}$ differs from the exact series: at $n = 0.7$--$0.8$ its third-order term projects $0.28$--$0.30$ onto the resummed result [metric (iii) of Appendix~\ref{app:h3}], against $0.60$--$1.05$ for the exact third order, and the agreement of Sec.~\ref{sec:mechanism} arises from compensation between orders.

With the measured $\chi_\mathrm{sp,ch}(\mathbf q, i\Omega_m)$ and vertices at $|m| \leq 3$ in Eq.~\eqref{eq:sigma_tpsc}, the spin term exceeds that of TPSC by a factor $1.28$--$1.34$ at $m=0$, falling to $0.97$--$1.26$ at $m=3$, and the resulting self-energy overshoots the local $\gamma$ by $36$--$53\%$ at $n \geq 0.7$ and $T/t \leq 0.14$, against $6$--$16\%$ for TPSC.
A constant fermion-boson vertex with $\lambda^2 = 0.65$--$0.73$, fixed by the local $\gamma$, gives $0.27$--$0.67$ of the CDet nonlocal scattering rate at $n = 0.6$--$0.9$, a mean of $0.47$ at $n = 0.6$--$0.85$, and $1.3$--$2.3$ times the CDet value at $n = 0.4$--$0.5$.
With the bare prefactor $U^2$, or in the FLEX normalization $U^2(\tfrac34\chi_\mathrm{sp} + \tfrac14\chi_\mathrm{ch} - \tfrac12\chi_0)$, the fractions are $0.36$--$0.79$ and $0.45$--$0.97$, and only at $(n=0.9, T/t=0.14)$ do they reach the CDet value within errors.
The nonlocal scattering rate of TPSC correlates with the CDet one at $0.68$--$0.89$ for $n = 0.7$--$0.9$ and $T/t \geq 0.10$ but has $0.38$--$0.88$ of its size, and for $n \geq 0.8$ and $T/t \leq 0.14$ its $c_1$ is $1.5$--$12\%$ larger in magnitude than the CDet value, at each point less than the CDet uncertainty ($3.5$--$42\%$).

\subsection{Fermi-surface anisotropy and frequency dependence}
\label{app:fs}

On the shell $|\varepsilon_\mathbf{k} - \mu_0| < t$ ($62$--$74$ irreducible momenta), with $O_h$ weights and all momenta at the common order, the variation of $Z_\mathrm{2pt}(\mathbf k)$ is its relative root-mean-square deviation from the shell mean, and the variation of the scattering rate at fixed band energy is the root-mean-square residual of a weighted linear or quadratic fit of $\gamma(\mathbf k)$ in $\varepsilon_\mathbf{k}$, relative to the shell mean.
Their noise distribution comes from $100$ parametric replicas per point, each with every coefficient redrawn within its Monte Carlo error and resummed again at the common order.
The quoted limit is the larger of the observed value and the one-sided $95\%$ upper bound on the noise-subtracted variation, recomputed also with each estimator alone.
The limits of Sec.~\ref{sec:structure} are those of the quadratic fit; the linear fit gives $0.97\%$, and a single estimator at most $1.1\%$ with the quadratic fit.
At $T/t=0.10$ and $0.05$ the common order is not converged ($|r| \geq 1$ or a rising next order, Appendix~\ref{app:sys}), and the same statistics bound only the truncated series.
At $(n=0.9, T/t=0.14)$ the variation predicted by $\Sigma_\mathrm{cl}$ is $1.17(15)\%$, with pattern correlation $0.89$.
For $n \leq 0.6$ the variation follows second order (correlation at least $0.96$) and amounts to $0.17$--$0.64$ of $\Delta\mathrm{Im}\Sigma_\mathrm{RMS}$.
The hot-spot comparison of Sec.~\ref{sec:structure} at $(n=0.8, T/t=0.14)$, where $\mathbf q^* = (\pi,\pi,0.7\pi)$ ($q_z^\mathrm{fit} = 0.76\pi$), compares the shell momenta whose partner $\mathbf k + \mathbf q^*$ lies within $t/4$ of the Fermi level with the rest of the shell.

\emph{Separability.}
The rank-1 fraction $s_1^2/\|A\|_F^2$, with $s_1$ the largest singular value of the $O_h$-weighted matrix $A = \mathrm{Re}\,M$ or $\mathrm{Im}\,M$, $M(\mathbf k,\omega_n) = \Sigma(\mathbf k,i\omega_n) - \Sigma_\mathrm{loc}(i\omega_n)$ over the first eight frequencies, is $0.96$--$0.998$ for the real and $0.94$--$0.994$ for the imaginary part at $U/t=4$ and $n = 0.7$--$0.8$, whereas noise-only surrogates of the same size give at most $0.66$.
The residual about the frequency average, relative to the norm of $A$, is $9$--$17\%$ for the real part at $T/t \leq 0.14$ but $25$--$89\%$ for the imaginary part, whose amplitude rises from $i\omega_0$.

\subsection{Scattering channels}
\label{app:h3}

\emph{Channel metrics.}
A piece $X$ of the nonlocal self-energy is characterized through its nonlocal part $dX$ ($X$ minus its $O_h$-weighted zone mean) and $\langle a,b\rangle = \sum_\mathbf{k} w_\mathbf{k} a_\mathbf{k} b_\mathbf{k}$, for the imaginary part at $i\omega_0$ with each piece multiplied by its $U^j$ at $U/t=4$.
For a piece $X$ of order $j$ and the resummed result $S$ the metrics are (i) the share of $\sigma_j$ by projection, $s_j(X) = \langle dX, d\sigma_j\rangle/\langle d\sigma_j, d\sigma_j\rangle$, the parts of one order adding to $1$; (ii) the correlation with the resummed result, $\mathrm{corr}(X) = \langle dX, dS\rangle/(|dX|\,|dS|)$; and (iii) the projection onto the resummed result, $\mathrm{proj}(X) = \langle dX, dS\rangle/\langle dS, dS\rangle$, all pieces, including the remainder $S - [U^2\sigma_2 + U^3\sigma_3 + U^4\sigma_4]$, adding to $1$.

\emph{Third order by channel.}
The bookkeeping of Appendix~\ref{app:norm} fixes this order: the Hartree-type terms vanish at fixed density, and the counterterm first enters an internal line at fourth order.
With $k = (\mathbf k, i\omega_n)$, $q = (\mathbf q, i\Omega_m)$ and $(T/N)\sum_q$ over both, the third-order coefficient therefore consists of exactly two diagrams built from $G_0$ at $\mu_0$, $\sigma_3 = \Sigma_\mathrm{pp} + \Sigma_\mathrm{ph}$, the particle-particle and the transverse (spin-flip) particle-hole ladder,
\begin{align}
    \Sigma_\mathrm{pp}(k) & = \frac{T}{N}\sum_q \Phi(q)^2\, G_0(q-k), \nonumber \\
    \Sigma_\mathrm{ph}(k) & = \frac{T}{N}\sum_q \Pi(q)^2\, G_0(k-q),
    \label{eq:sigma3}
\end{align}
where $\Phi$ is the particle-particle bubble and $\Pi = -\chi_0/2$ the particle-hole bubble of one spin.
Evaluated with the exact lattice bubbles, the two diagrams reproduce the measured third-order coefficient at all $286$ momenta and all exported frequencies in $14$ runs ($n = 0.4$--$0.9$, $\beta t = 7$, $10$, $20$): the nonlocal part at $i\omega_0$ and $i\omega_1$ agrees in size within $0.5\%$, with shape correlation $1.0000$, and the overall scale within about $1\%$.
By metric (i) above the particle-particle ladder is $68$--$84\%$ of $\sigma_3$ at $n = 0.7$--$0.9$ for $T/t = 0.05$--$0.14$.
Because a line of one spin couples only to the density of the other, the $U^3$ terms of the longitudinal spin and charge chains cancel, and $\Sigma_\mathrm{ph}$ is the whole $U^3$ content of bare-vertex (FLEX-type) particle-hole spin-fluctuation exchange.

\emph{Fourth order.}
The fourth-order coefficient of the density-fixed series consists of twelve tadpole-free topologies and the first counterterm on each of the three sunset lines.
Two groups are evaluated exactly at the seven $T/t = 0.14$ points.
The single-channel ladders are the spin-fluctuation chains, namely the longitudinal chain and the cubed spin-flip ladder, which agree numerically to within $10^{-13}$, and the cubed particle-particle ladder $P_4$.
The dressing is the sunset with one line dressed by the second-order self-energy and the counterterm.
The remainder, the measured $\sigma_4$ minus these two groups, consists of the vertex-correction diagrams: the four mixed particle-particle/particle-hole ladders, which include the $U^4$ terms of particle-particle-screened spin-fluctuation exchange, the crossed-bubble diagram and the sunset with a vertex-corrected bubble.
The vertex corrections, whose nonlocal part exceeds the Monte Carlo error of $\sigma_4$ by factors of $63$--$255$, are anti-collinear with the single-channel ladders at $n = 0.6$--$0.9$ (correlation $-0.98$ to $-0.999$, projection coefficient $-0.66$ to $-0.69$).
They cancel $66$--$69\%$ of the nonlocal part of the ladders but $86$--$90\%$ of their local part, so that the two groups together have $2.5$--$3.6$ times the nonlocal-to-local ratio of the ladders.
The chains correlate with the resummed nonlocal scattering rate at $0.90$--$0.95$ for $n = 0.6$--$0.9$, and the vertex corrections project to $-0.66$ and $-0.92$ at $n = 0.85$ and $0.9$.
At $n = 0.6$--$0.9$ the dressing projects to at most $0.07$ in magnitude, and the orders beyond the fourth to $0.17$, $0.20$ and $0.21$ at $n = 0.8$, $0.85$ and $0.9$.

\emph{Ladder series.}
The particle-particle ladder changes sign from third to fourth order at all seven fillings, with $|U^4/U^3| = 1.4$--$1.8$ in root mean square, because $U\Phi(\mathbf q, 0)$ reaches $1.6$--$1.9$ at $\mathbf q = 0$.
Resummed as a $T$-matrix on $G_0$ beyond second order, $(T/N)\sum_q U^3\Phi^2/(1+U\Phi)\,G_0(q-k)$, which is well defined because $1 + U\Phi$ has no zero for $U > 0$, it projects $+0.13$, $+0.18$, $+0.06$, $-0.01$ and $-0.07$ onto the resummed nonlocal scattering rate at $n = 0.6$, $0.7$, $0.8$, $0.85$ and $0.9$.
The bare-vertex spin-fluctuation self-energy on $G_0$, with the FLEX kernel of Appendix~\ref{app:models} and random-phase susceptibilities, overshoots the CDet nonlocal scattering rate by factors of $9$ and $11$ at $n = 0.6$ and $0.7$ and is undefined at $n \geq 0.8$ (Stoner pole).

\subsection{Single-site DMFT and the quasiparticle weight}
\label{app:dmft}
\label{app:z_freq}

The comparison of Sec.~\ref{sec:local} uses paramagnetic single-site DMFT on the lattice of the CDet runs: the local Green's function is the average over the $L=20$ momentum grid and the chemical potential fixes the density on that grid, so the finite-size effect of Appendix~\ref{app:finite_size} cancels in the comparison.
The impurity problem is solved by exact diagonalization with $N_b=6$ bath sites, the hybridization being fitted on the Matsubara axis up to $20t$ with weight $1/\omega_n$, and the loop is iterated until $\Sigma$ changes by less than $10^{-6}$.
Replacing $N_b=6$ by $N_b=5$ at all entries and by $N_b=7$ at three, or the hybridization-fit window of $20t$ by $10t$ or $40t$ at two, changes $Z$ and $\gamma$ by at most $0.02\%$.
The solver reproduces the second-order self-energy at weak coupling with a residual of order $U^3$, particle-hole symmetry at half filling, and the high-frequency moment $U^2(n/2)(1-n/2)$ of $\mathrm{Im}\,\Sigma$.
Only $\mathrm{Im}\,\Sigma$ is compared, through $Z_\mathrm{2pt}$ and $\gamma$, since the CDet $\mathrm{Re}\,\Sigma$ omits the Hartree constant $Un/2$ (Sec.~\ref{sec:CDet}).

Two local baselines on the same grid give the comparison its scale: the local part of the second-order term $U^2\sigma_{2,0}$, and second-order (IPT-type) DMFT, whose impurity self-energy is second order in the Hartree-shifted Weiss field, with $\mu$ refit to the density at every iteration.
At $n \leq 0.6$ DMFT is $5$--$70$ times closer to CDet than either baseline, and at $n = 0.85$--$0.9$ all three fall below the CDet $\gamma$, DMFT by $0.5$--$1.3$ and the baselines by $1.3$--$3.5$ standard deviations; these pulls include the enlarged truncation term of Appendix~\ref{app:sys}, which at $n \geq 0.8$ covers a $\gamma$ that still grows with order.
A commensurate N\'eel solution seeded at every point of Table~\ref{tab:dmft} decays to the paramagnet except at $(n=0.9, T/t=0.14)$ for $U/t=6$ and $8$ (marked $^\sharp$).

\emph{Quasiparticle weight.}
For the local self-energy in the low-frequency form of Sec.~\ref{sec:quantities}, $Z_1$ equals the Fermi-liquid $Z$, whereas $1/Z_\mathrm{2pt} = 1/Z_1 - 4\pi A T$: the secant between $\omega_0$ and $\omega_1$ carries a bias of order $T$, and a quadratic through the three lowest Matsubara frequencies, $Z_\mathrm{3pt}$, would equal $Z_1$ if that form held exactly.
In the data the three are ordered as $Z_1 < Z_\mathrm{3pt} < Z_\mathrm{2pt}$ at all $90$ combinations with $n \leq 0.8$, and at $n=0.8$, $U/t=4$ the difference $Z_\mathrm{2pt} - Z_1$ shrinks on cooling from $0.142$ at $T/t=0.33$ to $0.035$ at $T/t=0.05$.
Single-site DMFT shows the same difference ($0.029$--$0.086$ at $T/t=0.14$, against $0.028$--$0.089$ in CDet), so the comparison with CDet, which uses one estimator on both sides, is unaffected; with $Z_\mathrm{3pt}$ in place of $Z_\mathrm{2pt}$ (CDet frequency data for $n \leq 0.8$) the DMFT differences at $U/t=4$ are $-0.4$ to $+0.3\%$.
At the level of individual momenta, no positive low-frequency slope $b_\mathbf{k} = 1 - 1/Z_\mathrm{2pt}(\mathbf k)$ exceeds its propagated resummation budget at any of the $37\,466$ momentum-point pairs of the $131$ retained combinations; positive central slopes occur only at $n \geq 0.9$ and $U/t=8$, where the two-point inversion is poorly conditioned.

\bibliography{paper_A}

\end{document}